\documentclass[twocolumn,superscriptaddress,showpacs,amsmath,amssymb,aps]{revtex4-1}
\usepackage{graphicx}
\usepackage{color}
\usepackage{comment}

\newcommand{\diff}{\mathrm{d}}
\newcommand{\rme}{\mathrm{e}}
\newcommand{\ham}{\mathcal{H}}

\begin{document}
\title{Upper and Lower Critical Decay Exponents of Ising Ferromagnets with Long-range Interaction}
\author{Toshiki Horita}
\affiliation{Department of Applied Physics, University of Tokyo, Tokyo 113-8656, Japan}
\author{Hidemaro Suwa}
\affiliation{Department of Physics, University of Tokyo, Tokyo 113-0033, Japan}
\author{Synge Todo}
\affiliation{Department of Physics, University of Tokyo, Tokyo 113-0033, Japan}
\affiliation{Institute for Solid State Physics, University of Tokyo, Kashiwa 277-8581, Japan}

\date{\today}
\begin{abstract}
We investigate the universality class of the finite-temperature phase transition of the two-dimensional Ising model with the algebraically decaying ferromagnetic long-range interaction, $J_{ij} = |\vec{r}_i -\vec{r}_j|^{-(d+\sigma)}$, where $d$ (=2) is the dimension of the system and $\sigma$ the decay exponent, by means of the order-$N$ cluster-algorithm Monte Carlo method. In particular, we focus on the upper and lower critical decay exponents, the boundaries between the mean-field-universality, intermediate, and short-range-universality regimes. At the critical decay exponents, it is found that the standard Binder ratio of magnetization at the critical temperature exhibits the extremely slow convergence as a function of the system size. We propose more effective physical quantities, the combined Binder ratio and the self-combined Binder ratio, both of which cancel the leading finite-size corrections of the conventional Binder ratio. Utilizing these techniques, we clearly demonstrate that in two dimensions the lower and upper critical decay exponents are $\sigma = 1$ and 7/4, respectively, contrary to the recent Monte Carlo and the renormalization-group studies [M. Picco, arXiv:1207.1018; T. Blanchard, {\it et al.}, Europhys. Lett. {\bf 101}, 56003 (2013)].
\end{abstract}


\maketitle

\section{Introduction}\label{chap:intro}

A system with Long-range interaction can exhibit substantially
different physics from the corresponding system only with short-range
interaction~\cite{Dyson1969,FisherMN1972,Sak1973}. Many materials show non-trivial phenomena to which
long-range nature of the interactions, such as
the dipole-dipole interaction~\cite{KraemerNPTPKKRAGPPSKZKR2012} and the Ruderman-Kittel-Kasuya-Yoshida (RKKY) interaction~\cite{PrueserDBUPLW2014}, plays an essential role. The
competition between the short-range and the long-range interactions has
been studied in the cold-atom systems~\cite{LandigHDLMDE2016}.

One of the simplest and the most fundamental playgrounds for
long-range interaction is the Ising model with the algebraically decaying interaction:
\begin{equation}
 \ham = - \sum_{i<j}^N J_{ij} S_i S_j,
 \label{eqn:ham}
\end{equation}
where $S_i$ $(=\pm 1)$ is the Ising spin on the $i$-th site, $J_{ij}$ the coupling constant
between two spins ($S_i$ and $S_j$), and $N$ the total number of spins.  The summation in Eq.~(\ref{eqn:ham}) runs over all
spin pairs. The algebraically decaying
ferromagnetic long-range interaction is expressed as
 \begin{equation}\label{J}
  J_{ij} = \frac{1}{ | \vec{r}_i - \vec{r}_j |^{d + \sigma} },
 \end{equation}
where $d$ is the dimension of the system, and $\vec{r}_i$ is the
coordinate of the $i$-th site on the square lattice. The
decay exponent, $\sigma$, in Eq.~(\ref{J}) should be positive to ensure the
extensiveness of the energy, i.e., finite energy density in the
thermodynamic limit. Otherwise, one has to introduce an appropriate system-size-dependent normalization factor. The system with the algebraically decaying long-range interaction shows the richer critical phenomena than those with
only nearest-neighbor interaction.  For sufficiently small $\sigma$, the
finite-temperature phase transition is expected to belong to the mean-field universality class: in the limit of $d+\sigma \to 0$, the system becomes the fully connected model, or the Husimi-Temperley model~\cite{Husimi1953,Temperley1954}.
On the other hand, when $\sigma$ is sufficiently large, the nearest neighbor interaction dominates and the transition belongs to the short-range universality class.  In the
``intermediate regime'' between the mean-field
and the short-range limits, the critical exponents that characterize the
universality class vary continuously~\cite{FisherMN1972,Sak1973}.
For the $d$-dimensional system with the long-range interaction, this continuous change of the critical exponents between the short-range and the mean-field universalities can be interpreted as the continuous change of the effective dimension between $d$ and the upper critical dimension of the corresponding short range model.
Although a number of theoretical and numerical studies~\cite{FisherMN1972,Sak1973,vanEnter1982,LuijtenB2002,Picco2012,BlanchardPR2013,KatzgraberLY2009,BanosFMY2012,AngeliniPR2014,DefenuTC2015} have been conducted in order to interpret the intermediate regime as non-integral dimensions, precise identification between the decay exponent, $\sigma$, and the effective dimension has not been well established so far, in spite of the simple form of the Hamiltonian~(\ref{eqn:ham}).

The main difficulty in precise estimation of the critical decay
exponents is the strong (likely logarithmic) correction-to-scaling
near the boundaries~\cite{FisherMN1972,Brezin1982,AngeliniPR2014,BrezinPR2014}. In the present paper, we
propose new physical quantities, the combined Binder ratio and the
self-combined Binder ratio, both of which cancel the leading corrections of the conventional Binder ratio.
Using these combined Binder ratios, we investigate their universal ratio, the value at the critical point, as a function of the decay exponent ($\sigma$) precisely for the two-dimensional case. It is clarified that the lower and upper critical decay exponents are $\sigma=1$ and $7/4$, respectively.

The present paper is organized as follows: in the next section, the previous theoretical and numerical works on the
present model are briefly reviewed. In Sec.~\ref{sec:fukuitodo}, we introduce the
Fukui-Todo cluster-algorithm Monte Carlo method~\cite{FukuiT2009} that reduces the computational cost of simulating long-range
interacting systems to $O(N)$. In Sec.~\ref{sec:combin},
new physical quantities, the combined Binder ratio and the
self-combined Binder ratio, are proposed together with the demonstration for the
fully connected and the nearest neighbor models. In
Sec.~\ref{sec:results}, we show our results of Monte Carlo simulations
for the model~(\ref{eqn:ham}) with coupling
constant~(\ref{J}). Our study is concluded in
Sec.~\ref{sec:conclusion}. The details on the generalized Ewald-summation technique and the improved estimators are given, respectively, in
Appendices~\ref{subsec:ewald} and \ref{subsec:improvedestimator}.

\section{Summary of Previous Works}\label{chap:prev}
The Ising model with the long-range interaction~(\ref{J}) has two limits, the fully connected  Ising model [($d+\sigma) \rightarrow 0$] and the Ising model only with short-range interaction ($\sigma \rightarrow \infty$). The critical property of the phase transition in these limits is well established. The coupling constants of the former model are given by
\begin{equation}
 J_{ij} = \frac{1}{N}.
\end{equation}
On the other hand, those of the latter are
\begin{equation}
 J_{ij} = \begin{cases}
   1 & \text{if $i$ and $j$ are nearest neighbor,} \\
   0 & \text{otherwise.}
   \end{cases}
\end{equation}
In Table.~\ref{tab:MFandSR}, the critical exponents of these models are summarized.
Note that the critical exponents of the fully connected (mean-field) model depends on its dimension because its physical quantities scale with the total number of sites $N=L^d$ instead of system length $L$. The well known values $\nu=1/2$ and $\eta=0$ for the mean-field model are those at the upper critical dimension, i.e., $d=4$.

\begin{table}[tb]
 \begin{center}
 \caption{Critical exponents of the $d$-dimensional fully connected Ising model and the two-dimensional short-range Ising model~\cite{NishimoriO2011}.}
 \label{tab:MFandSR}
 \begin{tabular}{cccc} \hline
     & fully connected & short-range \\  \hline\hline
   $\alpha$ & \quad $0$ (discontinuous) &  $0$ (log)  \\
   $\beta$ &  $1/2$ &  $1/8$  \\
   $\gamma$ & $1$ & $7/4$ \\
   $\delta$ & $3$ & $15$ \\
   $\nu$ & $2/d$ & $1$ \\
   $\eta$ & $2-d/2$ & $1/4$ \\ \hline
 \end{tabular}
 \end{center}
\end{table}

A number of theoretical and numerical studies have been conducted mainly based on the renormalization-group (RG) argument and the Monte Carlo simulations.
For the O($n$) model with the long-range interaction, Fisher {\it et al.}~\cite{FisherMN1972}
performed the RG analysis.  The O($n$) model is a
generalization of the Ising model ($n=1$).  They found three different regimes, depending on
the value of the exponent $\sigma$ as listed in
Table~\ref{tab:universalityLRI@fisher}. In the
intermediate regime, the critical exponent of the correlation
function, $\eta$, varies linearly to $\sigma$.  The results of
their RG analysis has a flaw that $\eta$ at $\sigma=2$, on the
boundary between the intermediate regime and the short range regime,
is not determined uniquely.  The exponent $\eta$ becomes zero if one
approaches to $\sigma=2$ from the side of the intermediate regime, while the finite value ($\eta_{\rm sr}=1/4$) in the short range
regime; thus, $\eta$ changes discontinuously at $\sigma=2$.

The boundary between the intermediate and the short-range regimes was more carefully considered by Sak~\cite{Sak1973} taking into
account the higher-order terms in the RG calculations. In
Ref.~\citenum{Sak1973}, it was concluded that the boundary is $\sigma=2-\eta_{\rm sr}$ instead of $\sigma=2$
as listed in Table.~\ref{tab:universalityLRI@sak}. Then,
the exponent $\eta$ becomes a continuous function of $\sigma$.
Its derivative, nevertheless, is discontinuous at $\sigma=2-\eta_{\rm sr}$ (and also at $\sigma=1$).

Several theoretical studies have reported different conclusions. Most are based on the RG approach and the $ \epsilon $ expansion of the Landau-Ginzburg effective Hamiltonian where the propagator contains the $p^\sigma$ term in addition to the ordinary $p^2$ term. For example, van Enter~\cite{vanEnter1982} pointed out that the long-range perturbation is relevant for $2- \eta_{\rm sr} \le \sigma \le 2 $ in contradiction to the result obtained by Sak~\cite{Sak1973}.

\begin{table}[tb]
 \begin{center}
 \caption{Renormalization-group prediction for the critical exponent $\eta$ by Fisher {\it et al.}~\cite{FisherMN1972}.}
 \label{tab:universalityLRI@fisher}
 \begin{tabular}{lcc} \hline
   & $\sigma$ & $\eta$  \\  \hline\hline
   mean field regime & $\sigma < d/2$ & $ 1 $ \\
   intermediate regime & \quad $ d/2 < \sigma < 2 $ \quad & $ 2 - \sigma $ \\
   short range regime & $ 2 < \sigma $ & $\eta_{\rm sr}$ \\ \hline
  \end{tabular}
  \end{center}
 \end{table}

 \begin{table}[tb]
 \begin{center}
 \caption{Renormalization-group prediction for the critical exponent $\eta$ by Sak~\cite{Sak1973}.}
 \label{tab:universalityLRI@sak}
 \begin{tabular}{lcc} \hline
   & $\sigma$ & $\eta$  \\  \hline\hline
   mean field regime & $\sigma < d/2$ & $ 1 $ \\
   intermediate regime & \quad $ d/2 < \sigma < 2-\eta_{\rm sr} $ \quad & $ 2 - \sigma $ \\
   short range regime & $ 2-\eta_{\rm sr} < \sigma $ & $\eta_{\rm sr}$ \\ \hline
  \end{tabular}
  \end{center}
 \end{table}

In the meantime, the first numerical study of the Ising model with the long-range interaction was reported by Luijten and Bl\"{o}te~\cite{LuijtenB2002}. By means of the cluster-algorithm Monte Carlo method, they calculated the exponent $\eta$ for $d=2$ as a function of $\sigma$, concluding that $ \eta = 2 - \sigma $ up to $2 - \sigma = \eta_{\rm sr}$ and $\eta = \eta_{\rm sr} = 1/4$ for larger $\sigma$. This result seems consistent with the RG prediction by Sak~\cite{Sak1973}. Its error bar, however, is too large to exclude the possibility proposed by van Enter~\cite{vanEnter1982}.
Recently, Picco~\cite{Picco2012} showed the more precise Monte Carlo data than before and concluded that the exponent $\eta$ varies smoothly, connecting the intermediate regime and the short range regime, which disagrees with the previous RG analysis. Blanchard {\it et al.}~\cite{BlanchardPR2013} then supported the numerical result by means of the renormalization-group analysis with the double expansion. Nonetheless, another Monte Carlo calculation by Angelini {\it et al}.~\cite{AngeliniPR2014} and RG study by Defenu {\it et al}.~\cite{DefenuTC2015} agree with the result by Sak~\cite{Sak1973}. This discrepancy between the previous researches has been an enigma for years.

\section{Monte Carlo method for Systems with long-range interaction}\label{sec:fukuitodo}

The main difficulty in simulating the system with the long-range interaction is the large calculation cost. The number of pairs is $N(N-1)/2 \sim N^2$ for an $N$-spin system. A naive update in the Monte Carlo simulation will suffer from the $O(N^2)$ computational cost. To
make matters worse, the system with the long-range interaction is known for the egregious finite-size and boundary effects compared to the short-range model. Much larger systems thus need to be calculated, for the estimation of critical exponents, than the case of the short-range interaction. In addition, it becomes important to take into account interactions from mirror-image cells across the periodic boundaries and the Ewald summation to further reduce the finite-size effect.

A first efficient algorithm for the Ising model with the long-range interaction was proposed by Luijten and Bl\"ote~\cite{LuijtenB1995}, whose computational cost scales in $O(N \log N)$.
We will adopt the Fukui-Todo cluster algorithm~\cite{FukuiT2009} that is a more powerful approach; it further reduces the computational cost down to $O(N)$ for generic (unfrustrated) long-range interacting spin models~\cite{FukuiT2009,Tomita2009,TodoS2013}.

The Fukui-Todo cluster algorithm is based on the Swendsen-Wang cluster algorithm~\cite{SwendsenW1987}.
In the original Swendsen-Wang algorithm for inverse temperature $\beta$, we first activate each (interaction) bond with probability
\begin{equation}\label{validitySW}
  P_{ij} = 1- \exp(-2 \beta J_{ij}),
\end{equation}
when $S_i = S_j$. Then spin clusters, each of which consists of spins connected by activated bonds, are flipped independently to generate a next spin configuration.

In the Fukui-Todo cluster algorithm, first we introduce the extended
Fortuin-Kasteleyn representation~\cite{FortuinK1972,KawashimaG1995}.
The partition function of the Ising model~(\ref{eqn:ham}) is rewritten into the extended Fortuin-Kasteleyn representation:
 \begin{equation}
   Z = \sum_{\{ S_i \}} \sum_{\{ k_{ij} \}} \prod_{i<j}^{N} \Delta(\sigma_{ij},k_{ij}) V_{ij}(k_{ij}),
   \label{eqn:FK}
 \end{equation}
where $\sigma_{ij} = S_i S_j$ and $k_{ij}$ is a non-negative integer assigned to each spin pair $(i,j)$. The ``compatibility function'' $\Delta(\sigma_{ij}, k_{ij})$ and the ``weight'' $V_{ij}(k_{ij})$ in Eq.~(\ref{eqn:FK}) are defined as
\begin{equation}
  \Delta(\sigma_{ij},k_{ij}) = \begin{cases}
    1 & \text{if $\sigma_{ij} = 1$ or $k_{ij} = 0$}\\
    0 & \text{otherwise,}
   \end{cases}
 \end{equation}
 and
 \begin{equation}
  V_{ij}(k_{ij}) = \frac{\exp(-2 \beta J_{ij}) (2 \beta J_{ij})^{k_{ij}}}{{k_{ij}}!},
\end{equation}
respectively.
Due to the property of the compatibility function, antiparallel spin configuration ($\sigma_{ij} = -1$) is prohibited when $k_{ij} \ge 1$.  Then one can interpret a bond with $k_{ij} \ge 1$ and $k_{ij} = 0$ as an activated and deactivated bond, respectively, in a similar way to the original Swendsen-Wang algorithm.  Note that $V_{ij}(k_{ij})$ is the probability mass function of the Poisson distribution with mean $2 \beta J_{ij}$, and satisfies the normalization condition, $\sum_{k_{ij}=0}^{\infty} V_{ij}(k_{ij})=1$ for all $(i,j)$.

Based on the generalized Fortuin-Kasteleyn representation~(\ref{eqn:FK}), one Monte Carlo step of the Fukui-Todo cluster algorithm is performed as follows:
\begin{enumerate}
  \renewcommand{\labelenumi}{(\roman{enumi})}
\item Generate an integer $k$ from the Poisson distribution $f(k;\lambda_{\text{tot}})=\exp(-\lambda_{\text{tot}}) \lambda_{\text{tot}}^k/k!$, where $\lambda_{\text{tot}}=2 \beta \sum_{i<j}^N J_{ij}$.
\item Choose a pair $(i,j)$ according to the probability $\lambda_{ij}/\lambda_{\text{tot}}$, and increase $k_{i,j}$ by one if $\sigma_{ij} = 1$, where $\lambda_{ij}=2 \beta J_{ij}$.
  \item Repeat (ii) $k$ times in total.
  \item Spins connected by bonds with $k_{ij} > 0$ are considered to belong to the same cluster. Flip clusters at random and generate a new spin configuration as in the original Swendsen-Wang method.
\end{enumerate}
The probability that bond $(i,j)$ is activated can be expressed as
\begin{equation}\label{validityFT1}
  P_{ij} = \sum_{k=1}^{\infty} f(k;\lambda_{\rm tot}) \sum_{m=1}^{k} \frac{k!}{(k-m)!m!} \left(\frac{\lambda_{ij}}{\lambda_{\text{tot}}}\right)^m \left( 1- \frac{\lambda_{ij}}{\lambda_{\text{tot}}}\right)^{k-m},
\end{equation} 
since it is the probability that bond $(i,j)$ is chosen at least once. By using the normalization condition
\begin{equation}
  \sum_{m=0}^{k} \frac{k!}{(k-m)!m!} \left(\frac{\lambda_{ij}}{\lambda_{\text{tot}}}\right)^m \left( 1- \frac{\lambda_{ij}}{\lambda_{\text{tot}}}\right)^{k-m} = 1
\end{equation} 
of the binomial distribution, Eq.~(\ref{validityFT1}) can be rewritten as
\begin{equation}
  \begin{split}
    P_{ij} &= 1 - \sum_{k=0}^{\infty} f(k;\lambda_{\rm tot}) \left( 1- \frac{\lambda_{ij}}{\lambda_{\text{tot}}}\right)^{k} \\
     &= 1 - \exp(-\lambda_{\text{tot}})\sum_{k=0}^{\infty}\frac{1}{k!} (\lambda_{\text{tot}}-\lambda_{ij})^k \\
    &= 1 - \exp(-2 \beta J_{ij}).
  \end{split}
\end{equation}
That is, the activation probability in the Fukui-Todo method is equal to the one in the original Swendsen-Wang algorithm~(\ref{validitySW}). Thus the both stochastic processes are equivalent with each other, satisfying the detailed balance.

The computational cost of the Fukui-Todo method is proportional to the repeat count, $k$, of step~(ii). The average of $k$ is $\lambda_{\text{tot}}$, obeying the Poisson distribution, $f(k,\lambda_\text{tot})$. For a system with $N$ spins, $\lambda_{\text{tot}}$ is expressed as
\begin{equation}
  \label{eqn:lambda-tot}
  \begin{split}
    \lambda_{\text{tot}} &= 2 \beta \sum_{i<j}^{N} J_{ij} = \beta \sum_{i}^{N}\sum_{j \neq i}^N J_{ij} \\ 
    &\approx \beta \sum_{i}^N \int_{1}^{N^{1/d}} \!\!\! \diff r~ r^{d-1} J(r) \\ 
    &= \beta N \int_{1}^{N^{1/d}} \!\!\! \diff r~r^{d-1} J(r).
  \end{split}
\end{equation}
Here, we assume the translational invariance and that $J_{ij}$ depends only on the distance between spins. If $J(r)$ decays faster than $r^{-d}$, which is equivalent to the condition of the convergence of energy density for the ferromagnetic ordered state, the integral in the last line of Eq.~(\ref{eqn:lambda-tot}) converges to a finite value even in the thermodynamic limit. The convergence condition is unchanged even when the mirror-image cells are considered and the Ewald summation is taken (see Appendix.~\ref{subsec:ewald}). Thus, at a finite temperature, $\lambda_{\text{tot}}$ increases in proportional to $N$ instead of $N^2$ as long as $\sigma > 0$ and the energy density converges to a finite value. Because step~(ii) is done in $O(1)$ computational cost thanks to the Walker's method of alias (see the appendices of Ref.~\citenum{FukuiT2009}), the total cost of the one Monte Carlo step of the Fukui-Todo cluster algorithm is $O(\lambda_{\text{tot}}) \sim O(N)$.

\section{Combined Binder Ratio}\label{sec:combin}
In the next section, instead of the critical exponents, we use
the value of the Binder ratio for investigating the universality class of the
phase transition.  The the Binder ratio at the critical point, which is also referred to as the universal ratio, is constant and takes a universal value at the critical
point~\cite{LandauB2005}, since it is the ratio of two physical
quantities that have the same anomalous dimension at the critical
point. It can be usually calculated more accurately
than the critical exponents, which leads to more reliable identification of the
universality class~\cite{YasudaT2013}. One of the
simplest examples of such physical quantities is so-called the Binder
ratio~\cite{LandauB2005}
\begin{equation}
  Q = \frac{\langle m^2 \rangle^2}{\langle m^4 \rangle},
\end{equation}
where $m = \sum_i S_i$ and
$\langle \, \cdot \, \rangle$ denotes the Monte Carlo average.

\begin{figure}[tb]
 \begin{center}
  \includegraphics[width=8cm]{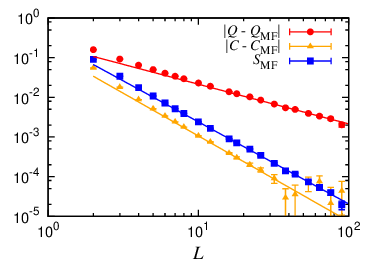}
  \caption{Convergence of the conventional Binder ratio (red circles), $Q$, the combined Binder ratio (orange triangles), $C$, and the self-combined Binder ratio (blue squares), $S_\text{MF}$ at the critical point for the fully connected model. The conventional Binder ratio converges to $Q_\text{MF}$ as $\sim L^{-1}$, while the combined Binder ratio and the self-combined Binder ratio converge as $\sim L^{-2}$ since the leading correction term is eliminated.}\label{fig:MF}
 \end{center}
\end{figure}

In practice, the Binder ratio exhibits some system-size dependence that is called the correction-to-scaling. We write the Binder ratio at the critical temperature as follows:
\begin{equation}
 Q(T_\text{c},L) = Q_\infty + f(L),
\end{equation}
where $Q_\infty$ denotes the universal ratio and $f(L)$ does the finite-size correction.
Note that the finite-size correction is not universal; i.e., it can be a different form even if the universality class is the same.
Although it is generally difficult to know the explicit form of $f(L)$, the expression of the leading correction term is known for the fully connected Ising model~\cite{LuijtenB1995}:
\begin{equation}
  \begin{split}
    Q(T_\text{c},N) & = \frac{\langle m^2 \rangle^2}{\langle m^4 \rangle} \\ 
    &= 4 \frac{\Gamma(3/4)^2}{\Gamma(1/4)^2} + \frac{16 \sqrt{3} \Gamma(3/4)^3}{5\Gamma(1/4)^3} N^{-\frac{1}{2}} + O(N^{-1}) \label{eq:binder24},
  \end{split}
\end{equation}
where $\Gamma(x)$ is the gamma function. In the case of the two-dimensional fully connected Ising model ($N=L^2$), the leading correction is proportional to $L^{-1}$. One can consider different combinations of the moment of magnetization to build various ``Binder ratios,'' whose universal ratio for the fully connected Ising model and its leading correction term at the critical point are also written explicitly as
\begin{equation}
 \frac{\langle m^2 \rangle^3}{\langle m^6 \rangle} \simeq \frac{4 \Gamma(3/4)^2}{3 \Gamma(1/4)^2} + \left( \frac{4 \sqrt{3} \Gamma(3/4)}{9 \Gamma(1/4)} - \frac{16 \sqrt{3} \Gamma(3/4)^3}{15 \Gamma(1/4)^3} \right)N^{-\frac{1}{2}},
\end{equation}
\begin{equation}
 \frac{\langle m^2 \rangle \langle m^4 \rangle}{\langle m^6 \rangle} \simeq \frac{1}{3} + \left( \frac{\sqrt{3} \Gamma(1/4)}{9 \Gamma(3/4)} - \frac{8\sqrt{3} \Gamma(3/4)}{15 \Gamma(1/4)} \right)N^{-\frac{1}{2}},
\end{equation}
\begin{equation}
 \frac{\langle m^2 \rangle^4}{\langle m^8 \rangle} \simeq \frac{16 \Gamma(3/4)^4}{5 \Gamma(1/4)^4} + \frac{1536\sqrt{3}\Gamma(3/4)^5}{125\Gamma(1/4)^5}N^{-\frac{1}{2}}.
\end{equation}
By using these Binder ratios, one can eliminate the lowest order of the correction-to-scaling. The most simplest way is taking an appropriate linear combination, e.g.,
\begin{eqnarray}
  C(T_\text{c},N) = \frac{\langle m^2 \rangle^2}{\langle m^4 \rangle} - a \frac{\langle m^2 \rangle^3}{\langle m^6 \rangle}
  = C_\text{MF}  + O(N^{-1}),
  \label{combined-binder}
\end{eqnarray}
where
\begin{equation}
a = \frac{ \frac{16 \sqrt{3} \Gamma(3/4)^3}{5\Gamma(1/4)^3}}{\left( \frac{4 \sqrt{3} \Gamma(3/4)}{9 \Gamma(1/4)} - \frac{16 \sqrt{3} \Gamma(3/4)^3}{15 \Gamma(1/4)^3} \right)}
  \label{combined-binder-coeff}
\end{equation}
and $C_\text{MF}=0.2843448$. Hereafter, we call $C(T,N)$ as the combined Binder ratio.

\begin{figure}[tb]
 \begin{center}
  \includegraphics[width=8.0cm]{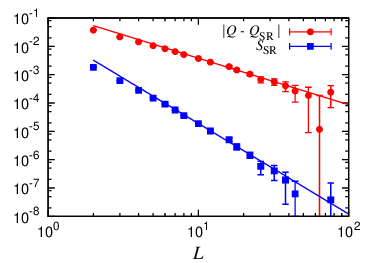}
  \caption{Convergence of the conventional Binder ratio (red circles), $Q$, and the self-combined Binder ratio (blue squares), $S_\text{SR}$ at the critical point for the nearest neighbor model. The conventional Binder ratio converges to $Q_\text{SR}$ as $\sim L^{-1.6}$, while the self-combined Binder ratio converges as $\sim L^{-3.2}$.}\label{fig:SR}
 \end{center}
\end{figure}

Unfortunately, the construction of the combined Binder ratio requires an explicit form of the leading correction term. The application to other universality classes rather than the mean-field universality is not practical. We then introduce another quantity that also has smaller finite-size corrections, the ``self-combined Binder ratio'':
\begin{equation}
 S(T,L) = \frac{1}{Q_\infty}Q(T,L) + Q_\infty \frac{1}{Q(T,L)} - 2 .
\end{equation}
This quantity is a linear combination of $Q$ and $Q^{-1}$. It is easily seen that regardless of the form of $f(L)$, leading correction of the universal ratio is reduced from $ O \left( f(L) \right) $ to $ O\left( f(L)^2 \right) $ if $Q_\infty$ is chosen as the exact universal ratio:
\begin{equation}
  \begin{split}
    S(T_{\rm c},L) &= \frac{1}{Q_\infty}Q(T_{\rm c},L) + Q_\infty \frac{1}{Q(T_{\rm c},L)} -2 \\
    &= \frac{ Q_\infty + f(L)}{Q_\infty} + \frac{1}{1 + \frac{1}{Q_\infty}f(L)} -2 \\
    &\simeq \frac{f(L)}{Q_\infty} -\frac{f(L)}{Q_\infty} +\left(\frac{1}{Q_\infty}f(L)\right)^2 \\
    &= O\left( f(L)^2 \right) 
  \end{split}
\end{equation}
In our analysis, we use the following self-combined Binder ratios:
\begin{equation}\label{eq:comMF}
 S_{\rm {MF}}(T,L) = \frac{1}{Q_{\rm {MF}}}Q(T,L) + Q_{\rm {MF}} \frac{1}{Q(T,L)} -2,
\end{equation}
\begin{equation}\label{eq:comSR}
 S_{\rm {SR}}(T,L) = \frac{1}{Q_{\rm {SR}}}Q(T,L) + Q_{\rm {SR}} \frac{1}{Q(T,L)} -2,
\end{equation}
where $Q_{\rm MF}=0.456947$ and $Q_{\rm {SR}}=0.856216$ are the universal ratio of the mean-field universality~\cite{LuijtenB1995} and short-range universality~\cite{KamieniarzB1993}, respectively.
The former, $S_{\rm {MF}}(T,L)$, converges to zero at the critical point for the mean-field universality class, and the latter, $S_{\rm {SR}}(T,L)$, does for the short-range universality class. Both should exhibit faster convergence than the conventional Binder ratio, $Q(T,L)$, since the leading correction is eliminated.

In Figs.~\ref{fig:MF} and \ref{fig:SR}, we compare the convergence of the conventional
Binder ratio, the combined Binder ratio, and the self-combined Binder
ratio to the limiting values for the fully connected model and the
nearest neighbor model in order to demonstrate the effectiveness of
our new quantities. For the both models, the (self-)combined Binder ratios converge much faster, in the double powers, than the conventional Binder ratio. Again note that the combined Binder ratio,
$C(T,L)$ is available only for the mean-field universality class and
not for the short-range case. We will apply these combined quantities to the phase transition of the Ising model with the algebraically decaying interaction.

\section{Results}\label{sec:results}

\begin{table}[bt]
 \begin{center}
 \caption{$\sigma$-dependence of the critical temperature, $T_{\rm c}$, and the normalized critical temperature, $\tilde{T}_{\rm c}$.  The exact results for the fully connected model (FC) and the nearest neighbor model (NN) are also included.}\label{tab:Tc}
  \begin{tabular}{lll} \hline
   $\sigma$ & $T_{\rm c}$ & $\tilde{T}_{\rm c}$ \\ \hline\hline
   0 (FC) & 1 & 1 \\
   0.6  &  12.555(1)   & 0.954033(81)\\
   0.8  &  9.76500(9)   & 0.923492(9) \\
   0.9  &  8.80870(32)  & 0.906553(33) \\
   1.0  &  8.03009(7)   & 0.888911(8) \\
   1.1  &  7.38232(41) \quad & 0.870847(49) \\
   1.2  &  6.83425(15)  & 0.852617(19) \\
   1.4  &  5.95819(12)  & 0.816679(16) \\
   1.6  &  5.29318(24)  & 0.782835(36) \\
   1.75 &  4.89500(12)  & 0.759570(19) \\
   1.9  &  4.56406(12)  & 0.738536(19) \\
   2.0  &  4.37427(14)  & 0.725805(24) \\
   $\infty$ (NN) \quad & 2.269185 & 0.5672963 \\ \hline
  \end{tabular}
 \end{center}
\end{table}

By means of the Fukui-Todo cluster method, the Monte Carlo simulation was performed on the two-dimensional $L \times L$ square lattice up to $L=4096$ for $\sigma = 0.8, 0.9, 1.0, \cdots, 1.75, 1.9$, and 2.0. The periodic boundary conditions were imposed and the effect of the mirror images was taken into account by the generalized Ewald summation (Appendix.~\ref{subsec:ewald}). The thermal averages of the
moments of the magnetization $\langle m^\alpha \rangle$
($\alpha=2,4,\cdots$) were calculated by the improved estimator
(Appendix~\ref{subsec:improvedestimator}).  
Measurement of physical quantities are performed for 16384 Monte Carlo steps, which is longer enough than the integrated autocorrelation time, e.g., $\tau_\text{int}=4.3$ and 4.0 for $\langle m^2 \rangle$ and $\langle m^4 \rangle$, respectively, at the critical point for $L=4096$ and $\sigma=1.75$. Before measurement, we discard 128 Monte Carlo steps. We have also confirmed that 128 Monte Carlo steps is longer enough than the exponential autocorrelation time, that is, there is no statistically significant difference between the measurement with 128 Monte Carlo thermalization steps and that with 256 Monte Carlo thermalization steps at the critical point for the largest system size.

\begin{figure}[tb]
 \begin{center}
  \includegraphics[width=8.0cm]{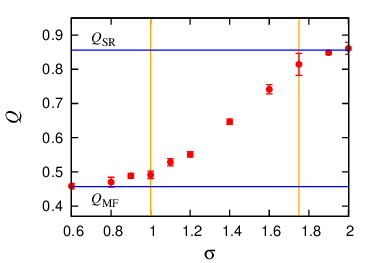}
  \caption{$\sigma$-dependence of the conventional Binder ratio at the critical point. The horizontal blue lines denote $Q_\text{MF}=0.456947$~\cite{LuijtenB1995} and $Q_\text{SR}=0.856216$~\cite{KamieniarzB1993}, respectively. The vertical orange lines denote the critical decay exponents $\sigma=1$ and $7/4$ predicted by Sak~\cite{Sak1973}.}\label{fig:binder}
 \end{center}
\end{figure}
\begin{figure}[tb]
 \begin{center}
  \includegraphics[width=8.0cm]{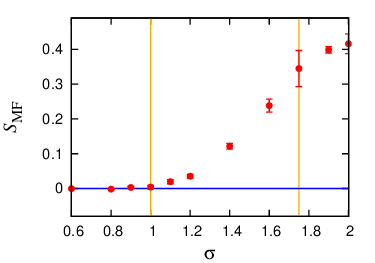}
  \caption{$\sigma$-dependence of the self-combined Binder ratio, $S_\text{MF}$, at the critical temperature. The vertical orange lines denote the critical decay exponents $\sigma=1$ and $7/4$ predicted by Sak~\cite{Sak1973}.}\label{fig:combinderMF}
 \end{center}
\end{figure}
\begin{figure}[tb]
 \begin{center}
  \includegraphics[width=8.0cm]{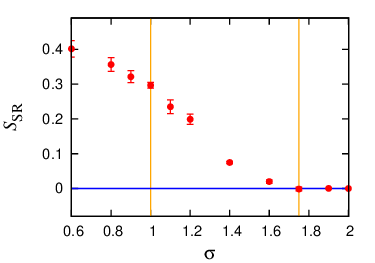}
  \caption{$\sigma$-dependence of the self-combined Binder ratio, $S_\text{SR}$, at the critical temperature. The vertical orange lines denote the critical decay exponents $\sigma=1$ and $7/4$ predicted by Sak~\cite{Sak1973}.}\label{fig:combinderSR}
 \end{center}
\end{figure}

\subsection{Critical temperature}
The critical temperature, $T_{\rm c}$, was estimated by the finite-size scaling of the Binder ratio.
First, we estimated the critical temperature for each system size, $T_{\rm c}(L)$, as the crossing point of the conventional Binder ratios of system size $L$ and $2L$ ($L=64$, 128, 256, 512, 1024, 2048). The crossing point is estimated by the Bayesian scaling analysis~\cite{Harada2011,BSAweb}. Then, we extrapolated the value in the thermodynamic limit by assuming the following form:
\begin{equation}
  \label{eqn:extrapolation}
  T_{\rm c}(L) = T_{\rm c} + a L^{-b} ,
\end{equation}
where the exponent $b$ is related with the correction-to-scaling eponent, $\omega$, as $1/\nu + \omega$. We observed that the value of $b$ takes larger value in the intermediate regime as reported in Ref.~\citenum{AngeliniPR2014}. Its precise estimation, however, is quite difficult even with the precision of the present simulation.

The critical temperature obtained for each $\sigma$ is
summarized in Table.~\ref{tab:Tc}.
We confirmed that the extrapolation by assuming Eq.~(\ref{eqn:extrapolation}) produces estimates for the critical temperatures that agree with the values for the largest system size within the error bar. The estimation from the self-combined Binder ratio (crossing point) is also consistent with the result in Table.~\ref{tab:Tc}. In the table, the normalized
critical temperature is also listed, which is defined as
$\tilde{T}_{\rm c} = T_{\rm c} / \tilde{J}$ with $\tilde{J} = \sum_{i
  \ne j} J_{ij}/N$, the sum of coupling constants connecting a single
site to all the other sites. One can see that the normalized critical
temperature, $\tilde{T}_{\rm c}$, increases slowly as $\sigma$
decreases. This shift of $\tilde{T}_{\rm c}$ can be interpreted as the
suppression of fluctuations due to the increase of the effective
dimension.

\subsection{Binder ratio at critical temperature}
Next, let us discuss the universality class of the phase transition by
using the Binder ratio at the critical temperature. As we have already
discussed in Sec.~\ref{chap:intro}, the three different regions are
expected: the mean-field, the intermediate, and the short-range
regime.  We will focus on the boundary, the critical decay exponent,
separating the mean-field and the intermediate regimes, and the
intermediate and the short-range regimes.

We calculated the Binder
ratio at the critical temperature (Table.~\ref{tab:Tc}) for each $\sigma$ and $L$, and
extrapolated the value in the thermodynamic limit by assuming the following
form:
\begin{equation}
 Q(T_\text{c},L) = Q_\infty + a L^{-b}
\end{equation}
for each $\sigma$, where $a$ and $b$ are $\sigma$-dependent fitting parameters.

In Fig.~\ref{fig:binder}, the universal ratio of the conventional
Binder ratio, $Q$, is plotted as a function of $\sigma$. The extrapolated values of $Q$ are consistent with
$Q_\text{MF}$ in the cases for $\sigma=0.6$ and 0.8, and consistent with $Q_\text{SR}$ for $\sigma=1.9$ and 2. Meanwhile, the values deviate from $Q_\text{MF}$ or $Q_\text{SR}$ even outside the region,
$1 < \sigma < 7/4$, which is similar to the Monte Carlo result by Picco~\cite{Picco2012} and in contradiction to the conclusion by Sak~\cite{Sak1973}. We will show
below, however, that this deviation and the smooth change are artifacts
due to strong corrections to scaling.

Next, we examine the behavior of the self-combined Binder ratio. In
Fig.~\ref{fig:combinderMF}, $S_\text{MF}$ [Eq.~(\ref{eq:comMF})] is
plotted as a function of $\sigma$, which has
smaller corrections if the transition belongs to the mean-field
universality class. In contrast to the conventional Binder ratio,
$S_\text{MF}$ becomes zero within the error bar for $\sigma \le 1$. Moreover, it is
observed that $S_\text{MF}$ grows as $\sim (\sigma-1)^2$ for $\sigma >
1$. Thus, we conclude that the transition belongs to the mean-field
universality for $\sigma \le 1$ and the conventional Binder ratio
changes linearly as $|Q-Q_\text{MF}| \sim (\sigma-1)$ for $\sigma >
1$; i.e., the critical decay exponent between the mean-field and the
intermediate regimes is one and $Q$ (probably
the critical exponents as well) is non-analytic at $\sigma=1$.

The boundary between the intermediate and the short-range regimes was
investigated precisely by the use of the self-combined Binder ratio,
$S_\text{SR}$ [Eq.~(\ref{eq:comSR})], as shown in Fig.~\ref{fig:combinderSR}.  In a similar way to the previous case,
$S_\text{SR}$ decreases as $\sim (7/4 - \sigma)^2$ for $\sigma < 7/4$
and becomes zero for $\sigma \ge 7/4$. Thus, we conclude that the critical
decay exponent is $\sigma = 7/4$.

\begin{figure}[tb]
 \begin{center}
  \includegraphics[width=8.0cm]{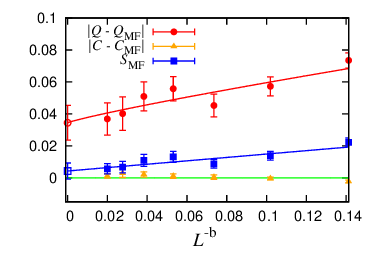}
  \caption{System-size dependence of the conventional Binder ratio ($Q$ with red circles), the combined Binder ratio ($C$ with orange triangles), and the self-combined Binder ratio ($S_\text{MF}$ with blue squares) at $\sigma=1$. Open symbols denote the extrapolated values to the thermodynamic limit by the fit to a function, $A + a L^{-b}$, where $A$, $a$, and $b$ are fitting parameters. The least-squares fitting yields $b \simeq 0.41$ and $0.47$ for $Q$ and $S_\text{MF}$, respectively. The $x$-axis is taken as $L^{-b}$ with $b=0.41$.}\label{fig:sigma1.0}
 \end{center}
\end{figure}

\begin{figure}[tb]
 \begin{center}
  \includegraphics[width=8.0cm]{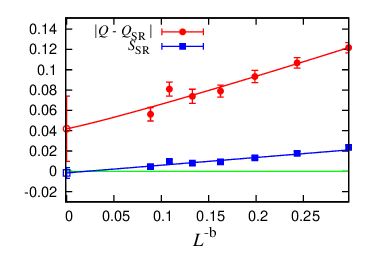}
  \caption{System-size dependence of the conventional Binder ratio ($Q$ with red circles) and the self-combined Binder ratio ($S_\text{SR}$ with blue squares) at $\sigma=7/4$. Open symbols denote the extrapolated values to the thermodynamic limit by the fit to a function, $A + a L^{-b}$, where $A$, $a$, and $b$ are fitting parameters. The least-squares fitting yields $b \simeq 0.32$ and $0.29$ for $Q$ and $S_\text{SR}$, respectively. The $x$-axis is taken as $L^{-b}$ with $b=0.32$.}\label{fig:sigma1.75}
 \end{center}
\end{figure}

The apparent deviation of the conventional Binder ratio from the
mean-field and short-range values at $\sigma=1$ and $7/4$,
respectively, is due to the existence of strong (likely logarithmic~\cite{FisherMN1972,AngeliniPR2014,BrezinPR2014})
corrections at the critical decay exponents. Indeed the nearest neighbor model exhibits the logarithmic
correction-to-scaling at the upper critical
dimension ($d=4$)~\cite{Brezin1982}. In Figs.~\ref{fig:sigma1.0} and
\ref{fig:sigma1.75}, we compare the system-size dependence of the
conventional Binder ratio, the combined Binder ratio (only for
$\sigma=1$), and the self-combined Binder ratio for $\sigma = 1$ and
$7/4$, respectively. We observe that in the both cases the combined or the self-combined Binder ratio converges to zero smoothly. On the other hand, the conventional Binder ratio has the stronger size dependence than the combined Binder ratios, which yields large error bars in the extrapolated values (open symbols). This observation indicates that the conventional Binder ratio presumably suffers from logarithmic-type corrections, while the (self-)combined Binder ratio is free from the strong corrections. In other words, we successfully remove the leading corrections of the conventional Binder ratio by considering the appropriate combination.
Note that the combined and the self-combined Binder ratios have smaller error bars than the conventional Binder ratio. It is expected that the statistical fluctuations of the terms of the combined quantities cancel with each other in the linear combination.

At last, it is quite interesting to see, in
Fig.~\ref{fig:sigma1.0}, that the combined Binder ratio
[Eq.~(\ref{combined-binder})] has even smaller corrections than the
self-combined Binder ratio at $\sigma=1$. The behavior of the conventional Binder
ratio (red circles in Fig.~\ref{fig:sigma1.0}) indicates that the
leading correction term might have a different exponent (smaller than
1/2) and have a quite different form from Eq.~(\ref{eq:binder24}). In
such a case, Eq.~(\ref{combined-binder}) with
coefficient~(\ref{combined-binder-coeff}) usually can not eliminate the
leading correction. The reason of
this accidental cancellation is not clear at the moment. It remains
as a future problem.


\section{Conclusion}\label{sec:conclusion}

We have performed the large-scale Monte Carlo simulations of the two-dimensional Ising model with the algebraically decaying long-range interaction and clarified the universality-regime boundaries that were in debate for decades.
For overcoming the difficulties that make the conventional analysis ineffective, we have utilized the several important techniques, the
$O(N)$ Fukui-Todo cluster algorithm (Sec.~\ref{sec:fukuitodo}), the (self-)combined
Binder ratio (Sec.~\ref{sec:combin}), the
generalized Ewald summation (Appendix.~\ref{subsec:ewald}), and
the generalization of the improved estimator for higher-order moments of magnetization
(Appendix.~\ref{subsec:improvedestimator}). We conclude that in two dimensions the
lower and upper critical decay exponents are $\sigma=1$ and $7/4$,
respectively, which is consistent with the renormalization-group prediction by Sak~\cite{Sak1973}, but contrary to the
recent Monte Carlo and renormalization studies by
Picco~\cite{Picco2012} and Blanchard {\it et al.}~\cite{BlanchardPR2013}.
The reason of the discrepancy is expected to be the strong (logarithmic)
correction-to-scaling effect near the critical decay exponents~\cite{BrezinPR2014}. We have observed that the logarithmic corrections hamper the standard analysis by the use of the conventional Binder ratio. Nevertheless, our (self-)combined Binder ratio removes the leading corrections of the conventional quantity and allows for the reliable identification of the universality class. 

The ``self-combined'' technique proposed in the present paper does not
require a concrete expression of the correction terms.  It should be
noted, however, that a precise value of the universal ratio in the
thermodynamic limit is needed beforehand to apply the proposed method
to identify the universality class of a given system.  Meanwhile, the
present ``self-combined'' technique is quite powerful to test whether a phase transition belongs to some well-known universality class or not, as
demonstrated in the present paper since the leading correction terms
are automatically eliminated as long as the correct limiting value is
used.  This property is expected to be useful not only for the
present case but also for a wide range of problems, especially for the
case with logarithmic corrections. The effectiveness of our approach
in other problems needs to be investigated in the future.

\section*{Acknowledgment}
The simulation code used in the present study has been developed based on the ALPS Library~\cite{ALPS2011, ALPSweb}.  The authors acknowledge the support by KAKENHI (No.\,23540438, 26400384, 16K17762) from JSPS, the HPCI System Research Project (No.\,hp140162), the HPCI Strategic Programs for Innovative Research (SPIRE) from MEXT, Japan, and the Computational Materials Science Initiative (CMSI).

\appendix
\section{Generalized Ewald Summation}\label{subsec:ewald}
In the simulation of systems with the long-range interaction under the periodic boundary conditions, we generally need to take into account interactions from mirror-image cells, which are imaginary systems across the periodic boundaries, in order to reduce strong finite-size corrections. The interaction between $S_i$ and $S_j$ is thus expressed as
\begin{equation}\label{eq:boundaryPotential}
  \ham_{ij} = -\sum_{\vec{\nu}}  \frac{1}{ | \vec{L}\cdot\vec{\nu} + \vec{r}_i - \vec{r}_j |^{d+\sigma} } S_i S_j,
\end{equation}
where $\vec{L} = (L_1,L_2,\cdots,L_d)$ denotes the size of the hypercubic system, and $\vec{\nu}=(\nu_1, \nu_2,\cdots,\nu_d)$ with $\nu_\alpha = 0, \pm 1, \pm 2, \cdots $ ($\alpha=1,2,\cdots,d$) represents the indices of image cells. In the following we consider only the case where $\sigma > 0$. The summation in Eq.~(\ref{eq:boundaryPotential}) are taken for all the image cells.
The simplest way to get an approximate value of
Eq.~(\ref{eq:boundaryPotential}) is truncating the summation up to some
threshold, $\nu_{\rm{max}}$, but this does not work well in practice, because
the summation of an algebraically decaying function converges very
slowly.
The Ewald-summation technique~\cite{Ewald1921,KarasawaG1989} is an efficient approach to calculate the contribution from image cells much faster~\cite{KretschmerB1979,Ueda1990}; the summation is broken into two parts, the
short-range term and the long-range term, and taken separately in the real space and the reciprocal space, respectively.
Although the Ewald summation is usually formulated only for integer exponents, $\sigma=0,1,2,\cdots$, such as the Coulomb interaction and the dipole interaction, here we generalize the Ewald summation for exponent of arbitrary positive real numbers.

Let us consider the potential of the following form:
 \begin{equation}\label{eq:rawEquation}
  \phi_m(i,j) = \sum_{\vec{\nu}} \frac{1}{|\vec{r}|^{m}},
 \end{equation}
where $m=d+\sigma$ and $\vec{r}=\vec{r}(\vec{\nu},\vec{r}_i,\vec{r}_j)\equiv\vec{L}\cdot\vec{\nu} + \vec{r}_i - \vec{r}_j$. We separate Eq.~(\ref{eq:rawEquation}) into two terms by using the gamma function:
 \begin{eqnarray}
  \phi_m(i,j) &=& \sum_{\vec{\nu}} \frac{1}{\Gamma(m/2)} \int_{0}^{\infty} \frac{1}{|\vec{r}|^m}t^{\frac{m}{2}-1}e^{-t} \diff t \nonumber \\
  &=& \sum_{\vec{\nu}} \Bigg[ \frac{1}{\Gamma(m/2)} \int_{(\kappa |\vec{r}|)^2}^{\infty} \frac{1}{|\vec{r}|^m}t^{\frac{m}{2}-1}e^{-t} \diff t \nonumber\\
  & & \qquad + \frac{2}{\Gamma(m/2)} \int_{0}^{\kappa} \rho^{m-1}e^{-|\vec{r}|^2 \rho^2} \diff  \rho \Bigg] \nonumber \\
  &=& \phi_m^{(1)}(i,j) + \phi_m^{(2)}(i,j).
 \end{eqnarray}
Here, $\Gamma(x)$ denotes the gamma function:
 \begin{equation}
  \Gamma(x) = \int_0^\infty t^{x -1} \rme^{-t} \diff t,
 \end{equation}
$\kappa$ is an arbitrary positive real number,
 \begin{equation}
   \label{eq:shortRangeTerm}
  \phi_m^{(1)}(i,j) = \sum_{\vec{\nu}} \frac{1}{\Gamma(m/2)} \frac{1}{|\vec{r}|^m} \int_{(\kappa |\vec{r}|)^2}^{\infty} t^{\frac{m}{2}-1}e^{-t} \diff t
 \end{equation}
represents the short-range term, and
 \begin{equation}\label{eq:derivationRealEwald2}
  \phi_m^{(2)}(i,j) = \sum_{\vec{\nu}} \frac{2}{\Gamma(m/2)} \int_{0}^{\kappa} \rho^{m-1}e^{-|\vec{r}|^2 \rho^2} \diff  \rho
 \end{equation}
does the long-range term. Note that in Eq.~(\ref{eq:derivationRealEwald2}) we have replaced the integration variable, $t$, by $(|\vec{r}| \rho)^2$.

Next, we transform Eq.~(\ref{eq:derivationRealEwald2}) from the integration in the real space into that in the reciprocal space. Let us consider the two-dimensional case, i.e., $d=2$, $\vec{L}=(L_x,L_y)$, $\nu=(\nu_x,\nu_y)$, $\vec{r}_i = (r_{ix},r_{iy})$, etc. We factor $ \sum_{\vec{\nu}} e^{-|\vec{r}|^2 \rho^2}$ in Eq.~(\ref{eq:derivationRealEwald2}) into the two components for each coordinate axis as
\begin{equation}\label{eq:derivationRealEwald2.5}
  G(\vec{r}_i,\vec{r}_j) = \sum_{\vec{\nu}} e^{-|\vec{r}|^2 \rho^2}
   = G_x(r_{ix},r_{jx}) G_y(r_{iy},r_{jx}),
 \end{equation}
where
 \begin{equation}\label{eq:derivationRealEwald3}
  G_x(r_{ix},r_{jx}) = \sum_{\nu_x=-\infty}^{\infty} \exp[-(L_x \nu_x + r_{ix} - r_{jx})^2 \rho^2] 
 \end{equation}
and
 \begin{equation}
  G_y(r_{iy},r_{jy}) = \sum_{\nu_y=-\infty}^{\infty} \exp[-(L_y \nu_y + r_{iy} - r_{jy})^2 \rho^2].
 \end{equation}
Then, Eq.~(\ref{eq:derivationRealEwald3}) is rewritten by taking the Fourier transform on $r_{ix}$ as
 \begin{equation}
   G_x(r_{ix},r_{jx}) = \sum_{h_x=-\infty}^{\infty} \frac{1}{L_x} A_x(h_x,r_{jx}) \exp\left(i \frac{2 \pi h_x}{L_x} r_{ix}\right),
   \label{eq:FourierG}
 \end{equation}
where
\begin{equation}
  \begin{split}
    A_x(h_x,r_{jx}) &= \int_0^{L_x} G_x(r_{ix},r_{jx}) \exp\left(-i \frac{2 \pi h_x}{L_x} r_{ix}\right) \diff r_{ix} \\
    &= \sum_{\nu_x} \int_{L_x\nu_x-r_{jx}}^{L_x(\nu_x+1) - r_{jx}} \exp\bigg[ -\xi^2 \rho^2\\ 
    & \qquad -i \frac{2 \pi h_x}{L_x}\xi +i 2 \pi h_x \nu_x -i \frac{2 \pi h_x}{L_x} r_{jx} \bigg] \diff \xi \\
    &= \int_{-\infty}^{\infty} \exp \bigg[-\xi^2 \rho^2 - i \frac{2 \pi h_x}{L_x} \xi - i \frac{2 \pi h_x}{L_x} r_{jx} \bigg] \diff \xi
   \label{eq:derivationRealEwald4}
  \end{split}
\end{equation}
with $ \xi = L_x \nu_x + r_{ix} - r_{jx}$. Note that $ \exp(i 2 \pi h_x \nu_x) $ vanishes because $ h_x \nu_x $ takes integral values.
Performing the Gaussian integration then yields
 \begin{equation}\label{eq:derivationRealEwald6}
  A_x(h_x,r_{jx}) = \frac{\sqrt{\pi}}{\rho} \exp\left( - \frac{\pi^2 h_x^2}{\rho^2 L_x^2} -i \frac{2 \pi h_x}{L_x} r_{jx} \right) .
\end{equation}
Next, by using Eq.~(\ref{eq:FourierG}) with Eq.~(\ref{eq:derivationRealEwald6}) and the similar expression for $G_y(r_{iy},r_{jy})$, Eq.~(\ref{eq:derivationRealEwald2.5}) is expressed as
 \begin{eqnarray}
  &G(\vec{r}_i,\vec{r}_j)& =
  \sum_{h_x,h_y} \frac{\pi}{L_x L_y \rho^2} \exp\Big[-\frac{\pi^2}{\rho^2} \left(\frac{h_x^2}{L_x^2} + \frac{h_y^2}{L_y^2}\right) \nonumber\\ 
  & & \mbox{} + i 2 \pi \left(\frac{h_x}{L_x}( r_{ix} - r_{jx} ) + \frac{h_y}{L_y} ( r_{iy} -r_{jy} ) \right)\Big]. \nonumber \\ \label{eq:derivationRealEwald7}
 \end{eqnarray}
Introducing $\vec{k} = \left(\frac{h_x}{L_x},\frac{h_y}{L_y}\right)$, Eq.~(\ref{eq:derivationRealEwald7}) is further simplified as
 \begin{eqnarray}
 G(\vec{r}_i,\vec{r}_j) &=& \sum_{\vec{k}} \frac{\pi}{L_x L_y \rho^2} \exp [-\frac{\pi^2}{\rho^2} |\vec{k}|^2 ] \nonumber \\ 
  & \times & \big(\cos [2 \pi \vec{k}\cdot(\vec{r}_i - \vec{r}_j) ]  + i \sin [2 \pi \vec{k}\cdot(\vec{r}_i - \vec{r}_j) ] \big). \nonumber \\ \label{eq:derivationRealEwald8}
\end{eqnarray}
Here, $ \exp [-\frac{\pi^2}{\rho^2} |\vec{k}|^2 ]$ and $\cos [2 \pi \vec{k}\cdot(\vec{r}_i - \vec{r}_j)]$ are even functions of $ \vec{k} $. On the other hand, $ \sin [2 \pi \vec{k}\cdot(\vec{r}_i - \vec{r}_j) ] $ is an odd function. Therefore, $ \sin [2 \pi \vec{k}\cdot(\vec{r}_i - \vec{r}_j) ] $ and $ \sin [2 \pi (-\vec{k})\cdot(\vec{r}_i - \vec{r}_j) ] $ are counterbalanced if summation is taken for all $\vec{k}$.

Extension of the above argument to arbitrary dimensions is straightforward. For the $d$-dimensional case, Eq(\ref{eq:derivationRealEwald8}) should be replace by
 \begin{eqnarray}
 G(\vec{r}_i,\vec{r}_j) &=& \frac{\pi^{d/2}}{V\rho^d} \sum_{\vec{k}} \exp [-\frac{\pi^2}{\rho^2} |\vec{k}|^2 ] \cos [2 \pi \vec{k}\cdot(\vec{r}_i - \vec{r}_j)], \nonumber \\
 \end{eqnarray}
where $ V = \Pi_{\alpha=1}^{d} L_\alpha $ is the volume of the system.
The long-range term~(\ref{eq:derivationRealEwald2}) is finally expressed as
\begin{equation}
  \begin{split}
    \phi_m^{(2)}(i,j) &= \frac{2 \pi^{d/2}}{\Gamma(m/2) V} \sum_{\vec{k}} \cos\left[2 \pi \vec{k}\cdot(\vec{r}_i - \vec{r}_j) \right] \\
   & \qquad \times \int_{0}^{\kappa}  \rho^{m-1-d} \exp\left[-\frac{\pi^2}{\rho^2} |\vec{k}|^2 \right] \diff \rho \\
   &= \frac{2 \pi^{d/2}}{\Gamma(m/2) V} \sum_{\vec{k}} \cos\left[2 \pi \vec{k}\cdot(\vec{r}_i - \vec{r}_j) \right] \\
    & \qquad \times \frac{1}{2} (\pi |\vec{k}|)^{m-d} \int_{\frac{\pi^2 |\vec{k}|^2}{\kappa^2}}^{\infty} t^{-\frac{1}{2}(m-d)-1 } \rme^{-t} \diff t .
    \label{eq:longRangeTerm}
  \end{split}
\end{equation}

In summary, the Ewald summation for generic exponent $\sigma$ is expressed as Eq.~(\ref{eq:rawEquation}) with the short-range term~(\ref{eq:shortRangeTerm}) and the long-range term~(\ref{eq:longRangeTerm}).
The terms in Eqs.~(\ref{eq:shortRangeTerm}) and (\ref{eq:longRangeTerm}) become small quite rapidly as $|\vec{\nu}|$ and $|\vec{k}|$ increase, respectively, due to the presence of the exponential functions. Note that the choice of the crossover parameter $\kappa$ affects the speed of convergence of the summations over $\vec{\nu}$ and $\vec{k}$ (or $\vec{h}$). We find empirically that $\kappa = 2/L$ is a reasonable choice, by which the both summations converge in double precision for $\max(|\nu_x|,|\nu_y|) \le 4$ and $\max(|h_x|,|h_y|) \le 4$.

The integrals in Eqs.~(\ref{eq:shortRangeTerm}) and (\ref{eq:longRangeTerm}) are calculated numerically. These integrals are known as the (upper) incomplete gamma function, which is defined as
 \begin{equation}\label{eq:incompleteGammaFunction}
  \Gamma(s,x) = \int_x^\infty t^{s-1} \rme^{-t} \diff t.
 \end{equation}
Eqs.~(\ref{eq:shortRangeTerm}) and (\ref{eq:longRangeTerm}) are expressed by the incomplete gamma function as
 \begin{equation}\label{eq:shortRangeTerm_igamma}
  \phi_m^{(1)}(i,j) = \sum_{\vec{\nu}} \frac{1}{\Gamma(m/2)} \frac{1}{|\vec{r}|^m} \Gamma \big(\frac{m}{2}, (\kappa |\vec{r}|)^2 \big) 
 \end{equation}
 \begin{eqnarray}
  &\displaystyle \phi_m^{(2)}(i,j) = \frac{2 \pi^{d/2}}{\Gamma(m/2) V} \sum_{\vec{k}} \cos\left[2 \pi \vec{k}\cdot(\vec{r}_i - \vec{r}_j) \right] \nonumber\\
   &\displaystyle \times \frac{1}{2} (\pi |\vec{k}|)^{m-d} \Gamma \big( -\frac{1}{2}(m-d),\frac{\pi^2 |\vec{k}|^2}{\kappa^2} \big), \label{eq:longRangeTerm_igamma}
 \end{eqnarray}
respectively.

In the present simulation, we calculated the incomplete gamma function numerically by the Boost C++ library~\cite{Boostweb}. The function~(\ref{eq:incompleteGammaFunction}), however, is defined only for positive real $s$ in the library. In the case for negative $s$, we used the following recursion formula obtained by the integration by parts:
\begin{eqnarray}
 \Gamma(s,x) &=& \int_x^\infty t^{s-1} \rme^{-t} \diff t \nonumber \\
    &=& \Big[ \frac{1}{s} t^s \rme^{-t} \Big]_{x}^{\infty} + \int_x^\infty \frac{1}{s} t^s \rme^{-t} \diff t \nonumber \\
    &=& -\frac{1}{s}x^s \rme^{-x} + \frac{1}{s} \Gamma(s+1,x) \nonumber \\
    &=& -\frac{1}{s}x^s \rme^{-x} - \frac{1}{s(s+1)}x^{s+1} \rme^{-x} \nonumber\\
    & & + \frac{1}{s(s+1)} \Gamma(s+2,x) \nonumber \\
    &\vdots& \nonumber \\ 
    &=& -\sum_{i=1}^{n} \Big\{ x^{s+i-1} \rme^{-x} \prod_{j=1}^{i} \big(\frac{1}{s+j-1}\big) \Big\} \nonumber\\
    & & + \prod_{i=1}^{n} \big(\frac{1}{s+i-1}\big) \Gamma(s+n,x), \label{eq:recursionFomula}
\end{eqnarray}
where $n=-\lfloor s \rfloor$. Note that this recursion formula breaks down when $\sigma$ is an even (positive) integer since $-(m-d)/2=-\sigma/2$ becomes a negative
integer. The denominator in Eq.~(\ref{eq:recursionFomula}) then becomes zero during the recursion.  For even $\sigma$, we stopped the recursion when the first argument of the incomplete gamma function becomes zero, and used
\begin{equation}
  \Gamma(0, x) = \int_x^\infty t^{-1} \rme^{-t} \diff t = -\mathrm{Ei}(-x),
\end{equation}
where $\mathrm{Ei}(-x)$ is the exponential integral.

\section{Improved Estimator}\label{subsec:improvedestimator}

In the cluster algorithm Monte Carlo methods, so-called the improved estimators for physical quantities are available, which are defined in terms of the cluster configuration instead of the spin configuration. Some improved estimators can drastically reduce the asymptotic variance of physical quantities because they take the average of a number of spin configurations generated from a cluster configuration automatically.  In the present work, we used the improved estimators for measuring the moments of magnetization, i.e., $\langle m^2 \rangle$, $\langle m^4 \rangle$, $\langle m^6 \rangle$, $\cdots$, to calculate the (combined) Binder ratios.

Let us consider a snapshot of graph configuration $g$ composed of $k$ clusters, and let $n_1, n_2,\cdots, n_k$ be the number of of spins included in each cluster. We define $m^\alpha(g)$ as an average of $m^\alpha$ over all possible $2^k$ different spin configurations that can be generated by flipping $k$ clusters independently:
\begin{equation}
  \begin{split}
  m^\alpha(g) &= \frac{1}{2^k} \sum_{s_1,\cdots,s_k} \Big(\sum_i^k n_i s_i \Big)^\alpha\\
   &= \frac{1}{2^k} \{ (+n_1 +n_2 +\cdots+n_k)^\alpha\\
   & \qquad \mbox{} +(-n_1 + n_2 + \cdots +n_k)^\alpha\\
   & \qquad \mbox{} +(+n_1 - n_2 + \cdots +n_k)^\alpha\\
   & \qquad \vdots \\
  & \qquad \mbox{} +(-n_1 - n_2 - \cdots -n_k)^\alpha \},\label{eq:m^xclu2}
  \end{split}
\end{equation}
where $s_i=\pm 1$ denotes the spin direction of the $i$-th cluster. In Eq.~(\ref{eq:m^xclu2}), terms including an odd power of $n_1,n_2,\cdots,n_k$ are counterbalanced by another terms with the opposite sign. Hence we only have to consider the terms with even powers for all $\{n_i\}$.
Thus, we can rewrite Eq.~(\ref{eq:m^xclu2}) as
\begin{eqnarray}
  m^\alpha(g) &=& \frac{1}{2^k} 2^k \left(n_1+n_2+\cdots+n_k \right)^\alpha_{\text{even terms}} \nonumber \\
  &=& \left(n_1+n_2+\cdots+n_k \right)^\alpha_{\text{even terms}}. \label{eq:m^xclu3}
\end{eqnarray}

Now we assume that $m^\alpha(g)$ can be expressed as
\begin{equation}\label{eq:improvedEstimator}
 m^\alpha(g) = \sum_p a_p \prod_{j \in p} C_j,
\end{equation}
with coefficients $\{a_p\}$ and
\begin{equation}
 C_j = \sum_i^k n_i^j.
\end{equation}
The summation $\sum_p$ in Eq.~(\ref{eq:improvedEstimator}) is taken over all possible partitions of $\alpha$ into a set of even integers $\{j\}$. We do not need, nevertheless, to consider the combinations with odd $j$ since they do not contribute after taking the average over the spin configurations as already mentioned.

We introduce a simple way to obtain $\{a_p\}$ by expanding the r.h.s.\ of Eqs.~(\ref{eq:m^xclu3}) and~(\ref{eq:improvedEstimator}), and comparing their coefficients.

\subsection{Second moment: $m^2(g)$}

In order to determine the improved estimator for the second moment of magnetization, it is enough to consider a graph configuration composed of only one cluster. In this case, Eq.~(\ref{eq:m^xclu3}) becomes
\begin{equation}\label{eq:m^2clu}
 m^2(g) = n_1^2
\end{equation}
On the other hand, Eq.~(\ref{eq:improvedEstimator}) is expressed as
\begin{equation}\label{eq:m^2imp}
 m^2(g) = a_1 \sum_i^k n_i^2 = a_1 n_1^2
.\end{equation}
By comparing Eqs.~(\ref{eq:m^2clu}) and (\ref{eq:m^2imp}), we obtain
\begin{equation}
 a_1 = 1
.\end{equation}
The improved estimator $m^2(g)$ is thus given by
\begin{equation}
 m^2(g) = C_2.
\end{equation}

\subsection{Fourth moment: $m^4(g)$}

Consider a graph configuration composed of two clusters.
From Eq.~(\ref{eq:m^xclu3}), we obtain
\begin{eqnarray}
 m^4(g) &=& (n_1 + n_2)^4_{\text{even terms}} \nonumber\\
   &=& (n_1^4+n_2^4) + 6 n_1^2 n_2^2 \label{eq:m^4clu}
\end{eqnarray}
On the other hand, Eq.~(\ref{eq:improvedEstimator}) is expressed as
\begin{eqnarray}
 m^4(g) &=& a_1 C_4 + a_2 C_2^2 \nonumber \\
 &=& (a_1 + a_2)(n_1^4+n_2^4) + 2 a_2 n_1^2 n_2^2 \label{eq:m^4imp}
.\end{eqnarray}
By comparing Eqs.~(\ref{eq:m^4clu}) and (\ref{eq:m^4imp}), we obtain
\begin{eqnarray}
 a_1 &=& -2 \nonumber\\
 a_2 &=& 3
.\end{eqnarray}
The improved estimator  $m^4(g)$ is thus given by
\begin{equation}
 m^4(g) = -2 C_4 +3 C_2^2.
\end{equation}

\subsection{Sixth moment: $m^6(g)$}

Consider a graph configuration composed of three clusters.
From Eq.~(\ref{eq:m^xclu3}), we obtain
\begin{eqnarray}
 m^6(g)&=& (n_1 + n_2 + n_3)^6_{\text{even terms}} \nonumber\\
   &=& (n_1^6+n_2^6+n_3^6) \nonumber\\ 
 & & \mbox{} + 15 \{n_1^4(n_2^2+n_3^2) + n_2^4(n_1^2+n_3^2) + n_3^4(n_1^2+n_2^2) \} \nonumber\\
   & & \mbox{} +90 n_1^2 n_2^2 n_3^2 \label{eq:m^6clu}
\end{eqnarray}
On the other hand, Eq.~(\ref{eq:improvedEstimator}) is expanded as
\begin{eqnarray}
 m^6(g) &=& a_1 C_6 + a_2 C_4 C_2 + a_3 C_2^3 \nonumber \\
   &=& (a_1 + a_2 +a_3)(n_1^6+n_2^6+n_3^6) \nonumber\\
   & & \mbox{} +(a_2+3a_3) \{ n_1^4(n_2^2+n_3^2)
 + n_2^4(n_1^2+n_3^2) \nonumber \\
 & & \qquad \mbox{} + n_3^4(n_1^2+n_2^2) \big\} \nonumber\\
   & & \mbox{} + 6 a_3 n_1^2 n_2^2 n_3^2 \label{eq:m^6imp}
.\end{eqnarray}
By comparing Eqs.~(\ref{eq:m^6clu}) and (\ref{eq:m^6imp}), we obtain
\begin{eqnarray}
 a_1 &=& 16 \nonumber\\
 a_2 &=& -30 \nonumber\\
 a_3 &=& 15.
\end{eqnarray}
The improved estimator $m^6(g)$ is thus given by
\begin{equation}
 m^6(g) = 16 C_6 -30 C_4 C_2 +15C_2^3
\end{equation}

\subsection{Eighth moment: $m^8(g)$}

Consider a graph configuration composed of four clusters.
From Eq.~(\ref{eq:m^xclu3}), we obtain
\begin{eqnarray}
  m^8(g) &=& (n_1 + n_2 + n_3 + n_4)^8_{\text{even terms}} \nonumber\\
  &=& \left( n_1^8+n_2^8+n_3^8+n_4^8 \right) \nonumber\\
   & & \mbox{} +28 \sum_i n_i^6 \sum_{j\neq i} n_j^2 \nonumber\\
   & & \mbox{} +420 \sum_i n_i^4 \sum_{j\neq i} n_j^2 \sum_{\ell\neq j, \ell\neq i} n_\ell^2 \nonumber\\
   & & \mbox{} +70 \sum_i n_i^4 \sum_{j\neq i} n_j^4 \nonumber\\
   & & \mbox{} +2520 n_1^2 n_2^2 n_3^2 n_4^2 \label{eq:m^8clu}
\end{eqnarray}
On the other hand, Eq.~(\ref{eq:improvedEstimator}) can be expanded as
\begin{eqnarray}
  m^8(g) &=& a_1 C_8 +a_2 C_6 C_2 +a_3 C_4 C_2^2 +a_4 C_4^2 +a_5 C_2^4 \nonumber\\
  &=& (a_1+a_2+a_3+a_4+a_5)\sum_i n_i^8 \nonumber\\
  & & \mbox{} + (a_2+2a_3+4a_5)\sum_i n_i^6 \sum_{j\neq i} n_j^2 \nonumber\\
  & & \mbox{} + (2a_3+12a_5)\sum_i n_i^4 \sum_{j\neq i} n_j^2 \sum_{\ell\neq j, \ell\neq i} n_\ell^2 \nonumber\\
  & & \mbox{} + (2a_3+2a_4+6a_5)\sum_i n_i^4 \sum_{j\neq i} n_j^4 \nonumber\\
  & & \mbox{} + 24a_5 n_1^2 n_2^2 n_3^2 n_4^2 \label{eq:m^8imp}
.\end{eqnarray}
By comparing Eqs.~(\ref{eq:m^8clu}) and (\ref{eq:m^8imp}), we obtain
\begin{eqnarray}
 a_1 &=& -272 \nonumber\\
 a_2 &=& 448 \nonumber\\
 a_3 &=& -420 \nonumber\\
 a_4 &=& 140 \nonumber\\
 a_5 &=& 105
.\end{eqnarray}
The improved estimator $m^8(g)$ is thus given by
\begin{equation}
 m^8(g) = -272 C_8 +448 C_6 C_2 -420 C_4 C_2^2 +140C_4^2 +105 C_2^4.
\end{equation}

\bibliography{main}

\begin{thebibliography}{38}%
\makeatletter
\providecommand \@ifxundefined [1]{%
 \@ifx{#1\undefined}
}%
\providecommand \@ifnum [1]{%
 \ifnum #1\expandafter \@firstoftwo
 \else \expandafter \@secondoftwo
 \fi
}%
\providecommand \@ifx [1]{%
 \ifx #1\expandafter \@firstoftwo
 \else \expandafter \@secondoftwo
 \fi
}%
\providecommand \natexlab [1]{#1}%
\providecommand \enquote  [1]{``#1''}%
\providecommand \bibnamefont  [1]{#1}%
\providecommand \bibfnamefont [1]{#1}%
\providecommand \citenamefont [1]{#1}%
\providecommand \href@noop [0]{\@secondoftwo}%
\providecommand \href [0]{\begingroup \@sanitize@url \@href}%
\providecommand \@href[1]{\@@startlink{#1}\@@href}%
\providecommand \@@href[1]{\endgroup#1\@@endlink}%
\providecommand \@sanitize@url [0]{\catcode `\\12\catcode `\$12\catcode
  `\&12\catcode `\#12\catcode `\^12\catcode `\_12\catcode `\%12\relax}%
\providecommand \@@startlink[1]{}%
\providecommand \@@endlink[0]{}%
\providecommand \url  [0]{\begingroup\@sanitize@url \@url }%
\providecommand \@url [1]{\endgroup\@href {#1}{\urlprefix }}%
\providecommand \urlprefix  [0]{URL }%
\providecommand \Eprint [0]{\href }%
\providecommand \doibase [0]{http://dx.doi.org/}%
\providecommand \selectlanguage [0]{\@gobble}%
\providecommand \bibinfo  [0]{\@secondoftwo}%
\providecommand \bibfield  [0]{\@secondoftwo}%
\providecommand \translation [1]{[#1]}%
\providecommand \BibitemOpen [0]{}%
\providecommand \bibitemStop [0]{}%
\providecommand \bibitemNoStop [0]{.\EOS\space}%
\providecommand \EOS [0]{\spacefactor3000\relax}%
\providecommand \BibitemShut  [1]{\csname bibitem#1\endcsname}%
\let\auto@bib@innerbib\@empty
\bibitem [{\citenamefont {Dyson}(1969)}]{Dyson1969}%
  \BibitemOpen
  \bibfield  {author} {\bibinfo {author} {\bibfnamefont {F.~J.}\ \bibnamefont
  {Dyson}},\ }\href@noop {} {\bibfield  {journal} {\bibinfo  {journal} {Commun.
  Math. Phys.}\ }\textbf {\bibinfo {volume} {12}},\ \bibinfo {pages} {91}
  (\bibinfo {year} {1969})}\BibitemShut {NoStop}%
\bibitem [{\citenamefont {Fisher}\ \emph {et~al.}(1972)\citenamefont {Fisher},
  \citenamefont {Ma},\ and\ \citenamefont {Nickel}}]{FisherMN1972}%
  \BibitemOpen
  \bibfield  {author} {\bibinfo {author} {\bibfnamefont {M.~E.}\ \bibnamefont
  {Fisher}}, \bibinfo {author} {\bibfnamefont {S.-K.}\ \bibnamefont {Ma}}, \
  and\ \bibinfo {author} {\bibfnamefont {B.~G.}\ \bibnamefont {Nickel}},\
  }\href@noop {} {\bibfield  {journal} {\bibinfo  {journal} {Phys. Rev. Lett.}\
  }\textbf {\bibinfo {volume} {29}},\ \bibinfo {pages} {917} (\bibinfo {year}
  {1972})}\BibitemShut {NoStop}%
\bibitem [{\citenamefont {Sak}(1973)}]{Sak1973}%
  \BibitemOpen
  \bibfield  {author} {\bibinfo {author} {\bibfnamefont {J.}~\bibnamefont
  {Sak}},\ }\href@noop {} {\bibfield  {journal} {\bibinfo  {journal} {Phys.
  Rev. B}\ }\textbf {\bibinfo {volume} {8}},\ \bibinfo {pages} {281} (\bibinfo
  {year} {1973})}\BibitemShut {NoStop}%
\bibitem [{\citenamefont {Kraemer}\ \emph {et~al.}(2012)\citenamefont
  {Kraemer}, \citenamefont {Nikseresht}, \citenamefont {Piatek}, \citenamefont
  {Tsyrulin}, \citenamefont {Piazza}, \citenamefont {Kiefer}, \citenamefont
  {Klemke}, \citenamefont {Rosenbaum}, \citenamefont {Aeppli}, \citenamefont
  {Gannarelli}, \citenamefont {Prokes}, \citenamefont {Podlesnyak},
  \citenamefont {Str{\"a}ssle}, \citenamefont {Keller}, \citenamefont
  {Zaharko}, \citenamefont {Kr{\"a}mer},\ and\ \citenamefont
  {R{\o}nnow}}]{KraemerNPTPKKRAGPPSKZKR2012}%
  \BibitemOpen
  \bibfield  {author} {\bibinfo {author} {\bibfnamefont {C.}~\bibnamefont
  {Kraemer}}, \bibinfo {author} {\bibfnamefont {N.}~\bibnamefont {Nikseresht}},
  \bibinfo {author} {\bibfnamefont {J.~O.}\ \bibnamefont {Piatek}}, \bibinfo
  {author} {\bibfnamefont {N.}~\bibnamefont {Tsyrulin}}, \bibinfo {author}
  {\bibfnamefont {B.~D.}\ \bibnamefont {Piazza}}, \bibinfo {author}
  {\bibfnamefont {K.}~\bibnamefont {Kiefer}}, \bibinfo {author} {\bibfnamefont
  {B.}~\bibnamefont {Klemke}}, \bibinfo {author} {\bibfnamefont {T.~F.}\
  \bibnamefont {Rosenbaum}}, \bibinfo {author} {\bibfnamefont {G.}~\bibnamefont
  {Aeppli}}, \bibinfo {author} {\bibfnamefont {C.}~\bibnamefont {Gannarelli}},
  \bibinfo {author} {\bibfnamefont {K.}~\bibnamefont {Prokes}}, \bibinfo
  {author} {\bibfnamefont {A.}~\bibnamefont {Podlesnyak}}, \bibinfo {author}
  {\bibfnamefont {T.}~\bibnamefont {Str{\"a}ssle}}, \bibinfo {author}
  {\bibfnamefont {L.}~\bibnamefont {Keller}}, \bibinfo {author} {\bibfnamefont
  {O.}~\bibnamefont {Zaharko}}, \bibinfo {author} {\bibfnamefont {K.~W.}\
  \bibnamefont {Kr{\"a}mer}}, \ and\ \bibinfo {author} {\bibfnamefont {H.~M.}\
  \bibnamefont {R{\o}nnow}},\ }\href@noop {} {\bibfield  {journal} {\bibinfo
  {journal} {Science}\ }\textbf {\bibinfo {volume} {336}},\ \bibinfo {pages}
  {1416} (\bibinfo {year} {2012})}\BibitemShut {NoStop}%
\bibitem [{\citenamefont {Prüser}\ \emph {et~al.}(2014)\citenamefont
  {Prüser}, \citenamefont {Dargel}, \citenamefont {Bouhassoune}, \citenamefont
  {Ulbrich}, \citenamefont {Pruschke}, \citenamefont {Lounis},\ and\
  \citenamefont {Wenderoth}}]{PrueserDBUPLW2014}%
  \BibitemOpen
  \bibfield  {author} {\bibinfo {author} {\bibfnamefont {H.}~\bibnamefont
  {Prüser}}, \bibinfo {author} {\bibfnamefont {P.~E.}\ \bibnamefont {Dargel}},
  \bibinfo {author} {\bibfnamefont {M.}~\bibnamefont {Bouhassoune}}, \bibinfo
  {author} {\bibfnamefont {R.~G.}\ \bibnamefont {Ulbrich}}, \bibinfo {author}
  {\bibfnamefont {T.}~\bibnamefont {Pruschke}}, \bibinfo {author}
  {\bibfnamefont {S.}~\bibnamefont {Lounis}}, \ and\ \bibinfo {author}
  {\bibfnamefont {M.}~\bibnamefont {Wenderoth}},\ }\href@noop {} {\bibfield
  {journal} {\bibinfo  {journal} {Nat. Commun.}\ }\textbf {\bibinfo {volume}
  {5}},\ \bibinfo {pages} {6417} (\bibinfo {year} {2014})}\BibitemShut
  {NoStop}%
\bibitem [{\citenamefont {Landig}\ \emph {et~al.}(2016)\citenamefont {Landig},
  \citenamefont {Hruby}, \citenamefont {Dogra}, \citenamefont {Landini},
  \citenamefont {Mottl}, \citenamefont {Donner},\ and\ \citenamefont
  {Esslinger}}]{LandigHDLMDE2016}%
  \BibitemOpen
  \bibfield  {author} {\bibinfo {author} {\bibfnamefont {R.}~\bibnamefont
  {Landig}}, \bibinfo {author} {\bibfnamefont {L.}~\bibnamefont {Hruby}},
  \bibinfo {author} {\bibfnamefont {N.}~\bibnamefont {Dogra}}, \bibinfo
  {author} {\bibfnamefont {M.}~\bibnamefont {Landini}}, \bibinfo {author}
  {\bibfnamefont {R.}~\bibnamefont {Mottl}}, \bibinfo {author} {\bibfnamefont
  {T.}~\bibnamefont {Donner}}, \ and\ \bibinfo {author} {\bibfnamefont
  {T.}~\bibnamefont {Esslinger}},\ }\href@noop {} {\bibfield  {journal}
  {\bibinfo  {journal} {Nature}\ }\textbf {\bibinfo {volume} {532}},\ \bibinfo
  {pages} {476} (\bibinfo {year} {2016})}\BibitemShut {NoStop}%
\bibitem [{\citenamefont {Husimi}(1953)}]{Husimi1953}%
  \BibitemOpen
  \bibfield  {author} {\bibinfo {author} {\bibfnamefont {K.}~\bibnamefont
  {Husimi}},\ }\href@noop {} {\bibfield  {journal} {\bibinfo  {journal} {Proc.
  Int. Conf. Theor. Phys.}\ ,\ \bibinfo {pages} {531}} (\bibinfo {year}
  {1953})}\BibitemShut {NoStop}%
\bibitem [{\citenamefont {Temperley}(1954)}]{Temperley1954}%
  \BibitemOpen
  \bibfield  {author} {\bibinfo {author} {\bibfnamefont {H.~N.~V.}\
  \bibnamefont {Temperley}},\ }\href@noop {} {\bibfield  {journal} {\bibinfo
  {journal} {Proc. Phys. Soc.}\ }\textbf {\bibinfo {volume} {67}},\ \bibinfo
  {pages} {233} (\bibinfo {year} {1954})}\BibitemShut {NoStop}%
\bibitem [{\citenamefont {van Enter}(1982)}]{vanEnter1982}%
  \BibitemOpen
  \bibfield  {author} {\bibinfo {author} {\bibfnamefont {A.~C.~D.}\
  \bibnamefont {van Enter}},\ }\href@noop {} {\bibfield  {journal} {\bibinfo
  {journal} {Phys. Rev. B}\ }\textbf {\bibinfo {volume} {26}},\ \bibinfo
  {pages} {1336} (\bibinfo {year} {1982})}\BibitemShut {NoStop}%
\bibitem [{\citenamefont {Luijten}\ and\ \citenamefont
  {Bl\"ote}(2002)}]{LuijtenB2002}%
  \BibitemOpen
  \bibfield  {author} {\bibinfo {author} {\bibfnamefont {E.}~\bibnamefont
  {Luijten}}\ and\ \bibinfo {author} {\bibfnamefont {H.~W.~J.}\ \bibnamefont
  {Bl\"ote}},\ }\href@noop {} {\bibfield  {journal} {\bibinfo  {journal} {Phys.
  Rev. Lett.}\ }\textbf {\bibinfo {volume} {89}},\ \bibinfo {pages} {025703}
  (\bibinfo {year} {2002})}\BibitemShut {NoStop}%
\bibitem [{\citenamefont {Picco}()}]{Picco2012}%
  \BibitemOpen
  \bibfield  {author} {\bibinfo {author} {\bibfnamefont {M.}~\bibnamefont
  {Picco}},\ }\href@noop {} {\ }\Eprint {http://arxiv.org/abs/arXiv:1207.1018}
  {arXiv:1207.1018} \BibitemShut {NoStop}%
\bibitem [{\citenamefont {Blanchard}\ \emph {et~al.}(2013)\citenamefont
  {Blanchard}, \citenamefont {Picco},\ and\ \citenamefont
  {Rajabpour}}]{BlanchardPR2013}%
  \BibitemOpen
  \bibfield  {author} {\bibinfo {author} {\bibfnamefont {T.}~\bibnamefont
  {Blanchard}}, \bibinfo {author} {\bibfnamefont {M.}~\bibnamefont {Picco}}, \
  and\ \bibinfo {author} {\bibfnamefont {M.~A.}\ \bibnamefont {Rajabpour}},\
  }\href@noop {} {\bibfield  {journal} {\bibinfo  {journal} {Europhys. Lett.}\
  }\textbf {\bibinfo {volume} {101}},\ \bibinfo {pages} {56003} (\bibinfo
  {year} {2013})}\BibitemShut {NoStop}%
\bibitem [{\citenamefont {Katzgraber}\ \emph {et~al.}(2009)\citenamefont
  {Katzgraber}, \citenamefont {Larson},\ and\ \citenamefont
  {Young}}]{KatzgraberLY2009}%
  \BibitemOpen
  \bibfield  {author} {\bibinfo {author} {\bibfnamefont {H.~G.}\ \bibnamefont
  {Katzgraber}}, \bibinfo {author} {\bibfnamefont {D.}~\bibnamefont {Larson}},
  \ and\ \bibinfo {author} {\bibfnamefont {A.~P.}\ \bibnamefont {Young}},\
  }\href@noop {} {\bibfield  {journal} {\bibinfo  {journal} {Phys. Lev. Lett}\
  }\textbf {\bibinfo {volume} {102}},\ \bibinfo {pages} {177205} (\bibinfo
  {year} {2009})}\BibitemShut {NoStop}%
\bibitem [{\citenamefont {Banos}\ \emph {et~al.}(2012)\citenamefont {Banos},
  \citenamefont {Fernandez}, \citenamefont {Martin-Mayor},\ and\ \citenamefont
  {Young}}]{BanosFMY2012}%
  \BibitemOpen
  \bibfield  {author} {\bibinfo {author} {\bibfnamefont {R.~A.}\ \bibnamefont
  {Banos}}, \bibinfo {author} {\bibfnamefont {L.~A.}\ \bibnamefont
  {Fernandez}}, \bibinfo {author} {\bibfnamefont {V.}~\bibnamefont
  {Martin-Mayor}}, \ and\ \bibinfo {author} {\bibfnamefont {A.~P.}\
  \bibnamefont {Young}},\ }\href@noop {} {\bibfield  {journal} {\bibinfo
  {journal} {Phys. Rev. B}\ }\textbf {\bibinfo {volume} {86}},\ \bibinfo
  {pages} {134416} (\bibinfo {year} {2012})}\BibitemShut {NoStop}%
\bibitem [{\citenamefont {Angelini}\ \emph {et~al.}(2014)\citenamefont
  {Angelini}, \citenamefont {Parisi},\ and\ \citenamefont
  {Ricci-Tersenghi}}]{AngeliniPR2014}%
  \BibitemOpen
  \bibfield  {author} {\bibinfo {author} {\bibfnamefont {M.~C.}\ \bibnamefont
  {Angelini}}, \bibinfo {author} {\bibfnamefont {G.}~\bibnamefont {Parisi}}, \
  and\ \bibinfo {author} {\bibfnamefont {F.}~\bibnamefont {Ricci-Tersenghi}},\
  }\href@noop {} {\bibfield  {journal} {\bibinfo  {journal} {Phys. Rev. E}\
  }\textbf {\bibinfo {volume} {89}},\ \bibinfo {pages} {062120} (\bibinfo
  {year} {2014})}\BibitemShut {NoStop}%
\bibitem [{\citenamefont {Defenu}\ \emph {et~al.}(2015)\citenamefont {Defenu},
  \citenamefont {Trombettoni},\ and\ \citenamefont {Codello}}]{DefenuTC2015}%
  \BibitemOpen
  \bibfield  {author} {\bibinfo {author} {\bibfnamefont {N.}~\bibnamefont
  {Defenu}}, \bibinfo {author} {\bibfnamefont {A.}~\bibnamefont {Trombettoni}},
  \ and\ \bibinfo {author} {\bibfnamefont {A.}~\bibnamefont {Codello}},\
  }\href@noop {} {\bibfield  {journal} {\bibinfo  {journal} {Phys. Rev. E}\
  }\textbf {\bibinfo {volume} {92}},\ \bibinfo {pages} {052113} (\bibinfo
  {year} {2015})}\BibitemShut {NoStop}%
\bibitem [{\citenamefont {Br\'ezin}(1982)}]{Brezin1982}%
  \BibitemOpen
  \bibfield  {author} {\bibinfo {author} {\bibfnamefont {E.}~\bibnamefont
  {Br\'ezin}},\ }\href@noop {} {\bibfield  {journal} {\bibinfo  {journal} {J.
  Phys. (France)}\ }\textbf {\bibinfo {volume} {43}},\ \bibinfo {pages} {15}
  (\bibinfo {year} {1982})}\BibitemShut {NoStop}%
\bibitem [{\citenamefont {Br\'ezin}\ \emph {et~al.}(2014)\citenamefont
  {Br\'ezin}, \citenamefont {Parisi},\ and\ \citenamefont
  {Ricci-Tersenghi}}]{BrezinPR2014}%
  \BibitemOpen
  \bibfield  {author} {\bibinfo {author} {\bibfnamefont {E.}~\bibnamefont
  {Br\'ezin}}, \bibinfo {author} {\bibfnamefont {G.}~\bibnamefont {Parisi}}, \
  and\ \bibinfo {author} {\bibfnamefont {F.}~\bibnamefont {Ricci-Tersenghi}},\
  }\href@noop {} {\bibfield  {journal} {\bibinfo  {journal} {J. Stat. Phys.}\
  }\textbf {\bibinfo {volume} {157}},\ \bibinfo {pages} {855} (\bibinfo {year}
  {2014})}\BibitemShut {NoStop}%
\bibitem [{\citenamefont {Fukui}\ and\ \citenamefont
  {Todo}(2009)}]{FukuiT2009}%
  \BibitemOpen
  \bibfield  {author} {\bibinfo {author} {\bibfnamefont {K.}~\bibnamefont
  {Fukui}}\ and\ \bibinfo {author} {\bibfnamefont {S.}~\bibnamefont {Todo}},\
  }\href@noop {} {\bibfield  {journal} {\bibinfo  {journal} {J. Comp. Phys.}\
  }\textbf {\bibinfo {volume} {228}},\ \bibinfo {pages} {2629} (\bibinfo {year}
  {2009})}\BibitemShut {NoStop}%
\bibitem [{\citenamefont {Nishimori}\ and\ \citenamefont
  {Ortiz}(2011)}]{NishimoriO2011}%
  \BibitemOpen
  \bibfield  {author} {\bibinfo {author} {\bibfnamefont {H.}~\bibnamefont
  {Nishimori}}\ and\ \bibinfo {author} {\bibfnamefont {G.}~\bibnamefont
  {Ortiz}},\ }\href@noop {} {\emph {\bibinfo {title} {Elements of Phase
  Transition and Critical Phenomena}}}\ (\bibinfo  {publisher} {Oxford
  University Press},\ \bibinfo {year} {2011})\BibitemShut {NoStop}%
\bibitem [{\citenamefont {Luijten}\ and\ \citenamefont
  {Bl\"ote}(1995)}]{LuijtenB1995}%
  \BibitemOpen
  \bibfield  {author} {\bibinfo {author} {\bibfnamefont {E.}~\bibnamefont
  {Luijten}}\ and\ \bibinfo {author} {\bibfnamefont {H.~W.~J.}\ \bibnamefont
  {Bl\"ote}},\ }\href@noop {} {\bibfield  {journal} {\bibinfo  {journal} {Int.
  J. Mod. Phys. C}\ }\textbf {\bibinfo {volume} {6}},\ \bibinfo {pages} {359}
  (\bibinfo {year} {1995})}\BibitemShut {NoStop}%
\bibitem [{\citenamefont {Tomita}(2009)}]{Tomita2009}%
  \BibitemOpen
  \bibfield  {author} {\bibinfo {author} {\bibfnamefont {Y.}~\bibnamefont
  {Tomita}},\ }\href@noop {} {\bibfield  {journal} {\bibinfo  {journal} {J.
  Phys. Soc. Jpn.}\ }\textbf {\bibinfo {volume} {78}},\ \bibinfo {pages}
  {014002} (\bibinfo {year} {2009})}\BibitemShut {NoStop}%
\bibitem [{\citenamefont {Todo}\ and\ \citenamefont {Suwa}(2013)}]{TodoS2013}%
  \BibitemOpen
  \bibfield  {author} {\bibinfo {author} {\bibfnamefont {S.}~\bibnamefont
  {Todo}}\ and\ \bibinfo {author} {\bibfnamefont {H.}~\bibnamefont {Suwa}},\
  }\href@noop {} {\bibfield  {journal} {\bibinfo  {journal} {J. Phys.: Conf.
  Ser.}\ }\textbf {\bibinfo {volume} {473}},\ \bibinfo {pages} {012013}
  (\bibinfo {year} {2013})}\BibitemShut {NoStop}%
\bibitem [{\citenamefont {Swendsen}\ and\ \citenamefont
  {Wang}(1987)}]{SwendsenW1987}%
  \BibitemOpen
  \bibfield  {author} {\bibinfo {author} {\bibfnamefont {R.~H.}\ \bibnamefont
  {Swendsen}}\ and\ \bibinfo {author} {\bibfnamefont {J.~S.}\ \bibnamefont
  {Wang}},\ }\href@noop {} {\bibfield  {journal} {\bibinfo  {journal} {Phys.
  Rev. Lett.}\ }\textbf {\bibinfo {volume} {58}},\ \bibinfo {pages} {86}
  (\bibinfo {year} {1987})}\BibitemShut {NoStop}%
\bibitem [{\citenamefont {Fortuin}\ and\ \citenamefont
  {Kasteleyn}(1972)}]{FortuinK1972}%
  \BibitemOpen
  \bibfield  {author} {\bibinfo {author} {\bibfnamefont {C.~M.}\ \bibnamefont
  {Fortuin}}\ and\ \bibinfo {author} {\bibfnamefont {P.~W.}\ \bibnamefont
  {Kasteleyn}},\ }\href@noop {} {\bibfield  {journal} {\bibinfo  {journal}
  {Physica}\ }\textbf {\bibinfo {volume} {57}},\ \bibinfo {pages} {536}
  (\bibinfo {year} {1972})}\BibitemShut {NoStop}%
\bibitem [{\citenamefont {Kawashima}\ and\ \citenamefont
  {Gubernatis}(1995)}]{KawashimaG1995}%
  \BibitemOpen
  \bibfield  {author} {\bibinfo {author} {\bibfnamefont {N.}~\bibnamefont
  {Kawashima}}\ and\ \bibinfo {author} {\bibfnamefont {J.~E.}\ \bibnamefont
  {Gubernatis}},\ }\href@noop {} {\bibfield  {journal} {\bibinfo  {journal} {J.
  Stat. Phys.}\ }\textbf {\bibinfo {volume} {80}},\ \bibinfo {pages} {169}
  (\bibinfo {year} {1995})}\BibitemShut {NoStop}%
\bibitem [{\citenamefont {Landau}\ and\ \citenamefont
  {Binder}(2005)}]{LandauB2005}%
  \BibitemOpen
  \bibfield  {author} {\bibinfo {author} {\bibfnamefont {D.~P.}\ \bibnamefont
  {Landau}}\ and\ \bibinfo {author} {\bibfnamefont {K.}~\bibnamefont
  {Binder}},\ }\href@noop {} {\emph {\bibinfo {title} {A Guide to {Monte}
  {Carlo} Simulations in Statistical Physics}}},\ \bibinfo {edition} {2nd}\
  ed.\ (\bibinfo  {publisher} {Cambridge University Press},\ \bibinfo {address}
  {Cambridge},\ \bibinfo {year} {2005})\BibitemShut {NoStop}%
\bibitem [{\citenamefont {Yasuda}\ and\ \citenamefont
  {Todo}(2013)}]{YasudaT2013}%
  \BibitemOpen
  \bibfield  {author} {\bibinfo {author} {\bibfnamefont {S.}~\bibnamefont
  {Yasuda}}\ and\ \bibinfo {author} {\bibfnamefont {S.}~\bibnamefont {Todo}},\
  }\href@noop {} {\bibfield  {journal} {\bibinfo  {journal} {Phys. Rev. E}\
  }\textbf {\bibinfo {volume} {88}},\ \bibinfo {pages} {061301(R)} (\bibinfo
  {year} {2013})}\BibitemShut {NoStop}%
\bibitem [{\citenamefont {Kamieniarz}\ and\ \citenamefont
  {Bl{\"o}te}(1993)}]{KamieniarzB1993}%
  \BibitemOpen
  \bibfield  {author} {\bibinfo {author} {\bibfnamefont {G.}~\bibnamefont
  {Kamieniarz}}\ and\ \bibinfo {author} {\bibfnamefont {H.~W.~J.}\ \bibnamefont
  {Bl{\"o}te}},\ }\href@noop {} {\bibfield  {journal} {\bibinfo  {journal} {J.
  Phys. A: Math. Gen.}\ }\textbf {\bibinfo {volume} {26}},\ \bibinfo {pages}
  {201} (\bibinfo {year} {1993})}\BibitemShut {NoStop}%
\bibitem [{\citenamefont {Harada}(2011)}]{Harada2011}%
  \BibitemOpen
  \bibfield  {author} {\bibinfo {author} {\bibfnamefont {K.}~\bibnamefont
  {Harada}},\ }\href@noop {} {\bibfield  {journal} {\bibinfo  {journal} {Phys.
  Rev. E}\ }\textbf {\bibinfo {volume} {84}},\ \bibinfo {pages} {056704}
  (\bibinfo {year} {2011})}\BibitemShut {NoStop}%
\bibitem [{BSA()}]{BSAweb}%
  \BibitemOpen
  \href@noop {} {\enquote {\bibinfo {title}
  {{http://kenjiharada.github.io/BSA/}},}\ }\BibitemShut {NoStop}%
\bibitem [{\citenamefont {Bauer}\ \emph {et~al.}(2011)\citenamefont {Bauer},
  \citenamefont {Carr}, \citenamefont {Evertz}, \citenamefont {Feiguin},
  \citenamefont {Freire}, \citenamefont {Fuchs}, \citenamefont {Gamper},
  \citenamefont {Gukelberger}, \citenamefont {Gull}, \citenamefont {Guertler},
  \citenamefont {Hehn}, \citenamefont {Igarashi}, \citenamefont {Isakov},
  \citenamefont {Koop}, \citenamefont {Ma}, \citenamefont {Mates},
  \citenamefont {Matsuo}, \citenamefont {Parcollet}, \citenamefont {Pawlowski},
  \citenamefont {Picon}, \citenamefont {Pollet}, \citenamefont {Santos},
  \citenamefont {Scarola}, \citenamefont {Schollw\"ock}, \citenamefont {Silva},
  \citenamefont {Surer}, \citenamefont {Todo}, \citenamefont {Trebst},
  \citenamefont {Troyer}, \citenamefont {Wall}, \citenamefont {Werner},\ and\
  \citenamefont {Wessel}}]{ALPS2011}%
  \BibitemOpen
  \bibfield  {author} {\bibinfo {author} {\bibfnamefont {B.}~\bibnamefont
  {Bauer}}, \bibinfo {author} {\bibfnamefont {L.~D.}\ \bibnamefont {Carr}},
  \bibinfo {author} {\bibfnamefont {H.~G.}\ \bibnamefont {Evertz}}, \bibinfo
  {author} {\bibfnamefont {A.}~\bibnamefont {Feiguin}}, \bibinfo {author}
  {\bibfnamefont {J.}~\bibnamefont {Freire}}, \bibinfo {author} {\bibfnamefont
  {S.}~\bibnamefont {Fuchs}}, \bibinfo {author} {\bibfnamefont
  {L.}~\bibnamefont {Gamper}}, \bibinfo {author} {\bibfnamefont
  {J.}~\bibnamefont {Gukelberger}}, \bibinfo {author} {\bibfnamefont
  {E.}~\bibnamefont {Gull}}, \bibinfo {author} {\bibfnamefont {S.}~\bibnamefont
  {Guertler}}, \bibinfo {author} {\bibfnamefont {A.}~\bibnamefont {Hehn}},
  \bibinfo {author} {\bibfnamefont {R.}~\bibnamefont {Igarashi}}, \bibinfo
  {author} {\bibfnamefont {S.~V.}\ \bibnamefont {Isakov}}, \bibinfo {author}
  {\bibfnamefont {D.}~\bibnamefont {Koop}}, \bibinfo {author} {\bibfnamefont
  {P.~N.}\ \bibnamefont {Ma}}, \bibinfo {author} {\bibfnamefont
  {P.}~\bibnamefont {Mates}}, \bibinfo {author} {\bibfnamefont
  {H.}~\bibnamefont {Matsuo}}, \bibinfo {author} {\bibfnamefont
  {O.}~\bibnamefont {Parcollet}}, \bibinfo {author} {\bibfnamefont
  {G.}~\bibnamefont {Pawlowski}}, \bibinfo {author} {\bibfnamefont {J.~D.}\
  \bibnamefont {Picon}}, \bibinfo {author} {\bibfnamefont {L.}~\bibnamefont
  {Pollet}}, \bibinfo {author} {\bibfnamefont {E.}~\bibnamefont {Santos}},
  \bibinfo {author} {\bibfnamefont {V.~W.}\ \bibnamefont {Scarola}}, \bibinfo
  {author} {\bibfnamefont {U.}~\bibnamefont {Schollw\"ock}}, \bibinfo {author}
  {\bibfnamefont {C.}~\bibnamefont {Silva}}, \bibinfo {author} {\bibfnamefont
  {B.}~\bibnamefont {Surer}}, \bibinfo {author} {\bibfnamefont
  {S.}~\bibnamefont {Todo}}, \bibinfo {author} {\bibfnamefont {S.}~\bibnamefont
  {Trebst}}, \bibinfo {author} {\bibfnamefont {M.}~\bibnamefont {Troyer}},
  \bibinfo {author} {\bibfnamefont {M.~L.}\ \bibnamefont {Wall}}, \bibinfo
  {author} {\bibfnamefont {P.}~\bibnamefont {Werner}}, \ and\ \bibinfo {author}
  {\bibfnamefont {S.}~\bibnamefont {Wessel}},\ }\href@noop {} {\bibfield
  {journal} {\bibinfo  {journal} {J. Stat. Mech.: Theo. Exp.}\ ,\ \bibinfo
  {pages} {P05001}} (\bibinfo {year} {2011})}\BibitemShut {NoStop}%
\bibitem [{ALP()}]{ALPSweb}%
  \BibitemOpen
  \href@noop {} {\enquote {\bibinfo {title} {{http://alps.comp-phys.org/}},}\
  }\BibitemShut {NoStop}%
\bibitem [{\citenamefont {Ewald}(1921)}]{Ewald1921}%
  \BibitemOpen
  \bibfield  {author} {\bibinfo {author} {\bibfnamefont {P.~P.}\ \bibnamefont
  {Ewald}},\ }\href@noop {} {\bibfield  {journal} {\bibinfo  {journal} {Ann.
  Phys}\ }\textbf {\bibinfo {volume} {64}},\ \bibinfo {pages} {253} (\bibinfo
  {year} {1921})}\BibitemShut {NoStop}%
\bibitem [{\citenamefont {Karasawa}\ and\ \citenamefont
  {Goddard}(1989)}]{KarasawaG1989}%
  \BibitemOpen
  \bibfield  {author} {\bibinfo {author} {\bibfnamefont {N.}~\bibnamefont
  {Karasawa}}\ and\ \bibinfo {author} {\bibfnamefont {W.~A.}\ \bibnamefont
  {Goddard}},\ }\href@noop {} {\bibfield  {journal} {\bibinfo  {journal} {J.
  Chem. Phys.}\ }\textbf {\bibinfo {volume} {93}},\ \bibinfo {pages} {7320}
  (\bibinfo {year} {1989})}\BibitemShut {NoStop}%
\bibitem [{\citenamefont {Kretschmer}\ and\ \citenamefont
  {Binder}(1979)}]{KretschmerB1979}%
  \BibitemOpen
  \bibfield  {author} {\bibinfo {author} {\bibfnamefont {R.}~\bibnamefont
  {Kretschmer}}\ and\ \bibinfo {author} {\bibfnamefont {K.}~\bibnamefont
  {Binder}},\ }\href@noop {} {\bibfield  {journal} {\bibinfo  {journal} {Z.
  Phys. B}\ }\textbf {\bibinfo {volume} {34}},\ \bibinfo {pages} {375}
  (\bibinfo {year} {1979})}\BibitemShut {NoStop}%
\bibitem [{\citenamefont {Ueda}(1990)}]{Ueda1990}%
  \BibitemOpen
  \bibfield  {author} {\bibinfo {author} {\bibfnamefont {A.}~\bibnamefont
  {Ueda}},\ }\href@noop {} {\emph {\bibinfo {title} {Computer Simulation (in
  Japanese)}}}\ (\bibinfo  {publisher} {Asakura},\ \bibinfo {address} {Tokyo},\
  \bibinfo {year} {1990})\BibitemShut {NoStop}%
\bibitem [{Boo()}]{Boostweb}%
  \BibitemOpen
  \href@noop {} {\enquote {\bibinfo {title} {{http://www.boost.org/}},}\
  }\BibitemShut {NoStop}%
\end{thebibliography}%

\end{document}